\documentclass[twocolumn]{aastex631}
\usepackage{amsmath,amssymb}
\usepackage{natbib}
\usepackage{graphicx}
\usepackage{xcolor}
\usepackage{ulem}
\usepackage[figuresright]{rotating}
\newcommand{\OII}{[O\,{\sc ii}]}
\newcommand{\OIII}{[O\,{\sc iii}]}
\newcommand{\Ha}{H$\alpha$}
\newcommand{\Hb}{H$\beta$}
\newcommand{\Hg}{H$\gamma$}
\newcommand{\Hd}{H$\delta$}
\newcommand{\NII}{[N\,{\sc ii}]}

\newcommand{\Lya}{Ly$\alpha$}
\newcommand{\Oabundance}{$12+\log ({\rm O/H})$}
\newcommand{\Msun}{$M_{\odot}$}
\newcommand{\Zsun}{$Z_{\odot}$}

\DeclareRobustCommand{\Erase}{\bgroup\markoverwith{\textcolor{red}{\rule[0.5ex]{2pt}{0.4pt}}}\ULon}

\newcommand{\SOKENDAI}{Department of Astronomical Science, The Graduate University for Advanced Studies, SOKENDAI, 2-21-1 Osawa, Mitaka, Tokyo, 181-8588, Japan}
\newcommand{\NAOJ}{National Astronomical Observatory of Japan, 2-21-1 Osawa, Mitaka, Tokyo, 181-8588, Japan}
\newcommand{\ICRR}{Institute for Cosmic Ray Research, The University of Tokyo, 5-1-5 Kashiwa-no-Ha, Kashiwa, Chiba, 277-8582, Japan}
\newcommand{\UTA}{Department of Astronomy, Graduate School of Science, The University of Tokyo, 7-3-1 Hongo, Bunkyo, Tokyo 113-0033, Japan}
\newcommand{\UTP}{Department of Physics, Graduate School of Science, The University of Tokyo, 7-3-1 Hongo, Bunkyo, Tokyo 113-0033, Japan}
\newcommand{\SUBARU}{Subaru Telescope, National Astronomical Observatory of Japan, National Institutes of Natural Sciences (NINS), 650 North A’ohoku Place, Hilo, HI 96720, USA}
\newcommand{\TSUKUBA}{Center for Computational Sciences, University of Tsukuba, Ten-nodai, 1-1-1 Tsukuba, Ibaraki 305-8577, Japan}

\accepted{for publication in ApJ on 19 April, 2023}
\shorttitle{EMPRESS XI}
\shortauthors{Nishigaki et al.}
\graphicspath{{./}{figures/}}

\begin{document}

\title{EMPRESS. XI.\\
SDSS and JWST Search for Local and $z\sim$ 4--5 Extremely Metal-Poor Galaxies (EMPGs):\\
Clustering and Chemical Properties of Local EMPGs
}

\author[0000-0003-4321-0975]{Moka Nishigaki}
\affiliation{\SOKENDAI}
\affiliation{\NAOJ}
\author[0000-0002-1049-6658]{Masami Ouchi}
\affiliation{\NAOJ}\affiliation{\ICRR}\affiliation{\SOKENDAI}
\affiliation{Kavli Institute for the Physics and Mathematics of the Universe (WPI), University of Tokyo, Kashiwa, Chiba 277-8583, Japan}
\author[0000-0003-2965-5070]{Kimihiko Nakajima}
\affiliation{\NAOJ}
\author[0000-0001-9011-7605]{Yoshiaki Ono}
\affiliation{\ICRR}
\author[0000-0002-1690-3488t]{Michael Rauch}
\affiliation{Observatories of the Carnegie Institution for Science, 813 Santa Barbara Street, Pasadena, CA 91101, USA}
\author[0000-0001-7730-8634]{Yuki Isobe}
\affiliation{\ICRR}\affiliation{\UTP}
\author[0000-0002-6047-430X]{Yuichi Harikane}
\affiliation{\ICRR}
\author{Kanako Narita}
\affiliation{\UTA}
\author[0000-0001-7869-2551]{Fakhri Zahedy}
\affiliation{Observatories of the Carnegie Institution for Science, 813 Santa Barbara Street, Pasadena, CA 91101, USA}
\author[0000-0002-5768-8235]{Yi Xu}
\affiliation{\ICRR}\affiliation{\UTA}
\author[0000-0002-1319-3433]{Hidenobu Yajima}
\affiliation{\TSUKUBA}
\author[0000-0002-0547-3208]{Hajime Fukushima}
\affiliation{\TSUKUBA}
\author[0000-0002-5661-033X]{Yutaka Hirai}
\affiliation{Department of Physics and Astronomy, University of Notre Dame, 225 Nieuwland Science Hall, Notre Dame, IN 46556, USA}
\affiliation{Astronomical Institute, Tohoku University, 6-3 Aoba, Aramaki, Aoba-ku, Sendai, Miyagi 980-8578, Japan}
\author[0000-0002-1418-3309]{Ji Hoon Kim}
\affiliation{Astronomy Program, Department of Physics and Astronomy, Seoul National University, 1 Gwanak-ro, Gwanak-gu, Seoul 08826, Republic of Korea}
\affiliation{SNU Astronomy Research Center, Seoul National University, 1 Gwanak-ro, Gwanak-gu, Seoul 08826, Republic of Korea}
\author[0000-0001-8819-6877]{Shigeki Inoue}
\affiliation{Faculty of Science, Hokkaido University, Sapporo, Hokkaido 060-0810, Japan}
\author[0000-0002-3801-434X]{Haruka Kusakabe}
\affiliation{Observatoire de Genève, Université de Genève, 51 Chemin de Pégase, 1290 Versoix, Switzerland}
\author[0000-0003-1700-5740]{Chien-Hsiu Lee}
\affiliation{W. M. Keck Observatory, 65-1120 Mamalahoa Hwy, Kamuela, HI 96743, USA}
\author[0000-0002-7402-5441]{Tohru Nagao}
\affiliation{Research Center for Space and Cosmic Evolution, Ehime University, Matsuyama, Ehime 790-8577, Japan}
\author[0000-0003-3228-7264]{Masato Onodera}
\affiliation{\SOKENDAI}\affiliation{\SUBARU}

\begin{abstract}
We search for local extremely metal-poor galaxies (EMPGs), selecting photometric candidates by broadband color excess and machine-learning techniques with the SDSS photometric data. 
After removing stellar contaminants
by shallow spectroscopy with Seimei and Nayuta telescopes,
we confirm that three candidates are EMPGs with 
0.05--0.1 $Z_\odot$
by deep Magellan/MagE spectroscopy for faint {\sc[Oiii]}$\lambda$4363 lines.
Using a statistical sample consisting of 105 spectroscopically-confirmed EMPGs taken from our study and the literature, we calculate cross-correlation function (CCF) of the EMPGs and all SDSS galaxies to quantify environments of EMPGs. Comparing another CCF of all SDSS galaxies and comparison SDSS galaxies in the same stellar mass range ($10^{7.0}-10^{8.4} M_\odot$), we find no significant ($>1\sigma$) difference between these two CCFs. We also compare mass-metallicity relations (MZRs) of the EMPGs and those of galaxies at $z\sim$ 0--4 
with a steady 
chemical evolution model
and find that the EMPG MZR is comparable with the model prediction on average.
These clustering and chemical properties of EMPGs are explained by a scenario of stochastic metal-poor gas accretion on metal-rich galaxies showing metal-poor star formation.
Extending the broadband color-excess technique to a high-$z$ EMPG search, we select 17 candidates of $z\sim$ 4--5 EMPGs with the deep ($\simeq30$ mag) near-infrared JWST/NIRCam images obtained by ERO and ERS programs. We find galaxy candidates with negligible {\sc[Oiii]}$\lambda\lambda$4959,5007 emission weaker than the local EMPGs and known high-$z$ galaxies, suggesting that some of these candidates may fall in 0--0.01 $Z_\odot$, which potentially break the lowest metallicity limit known to date.
\end{abstract}


\defcitealias{Kojima2020}{Paper~I}
\defcitealias{SA16}{SA16}
\defcitealias{Isobe2021}{Paper~III}
\defcitealias{Isobe2022a}{Paper~IV}
\defcitealias{Nakajima2022}{Paper~V}
\defcitealias{Xu2022}{Paper~VI}
\defcitealias{Isobe2022b}{Paper~IX}

\section{Introduction} \label{sec:intro}
Extremely metal-poor galaxies (EMPGs) in the local universe are important to understand the early phase of galaxy formation.
EMPGs are defined as galaxies with a gas-phase metallicity
of \Oabundance\ $< 7.69$ 
\citep[i.e., $Z <0.1\ Z_\odot$;][hereafter Paper I]{Kojima2020}.
The range where $Z <0.01$ \Zsun\ corresponds
to the chemical evolutionary stage with 
no dominant effects of Type Ia supernovae (SNe) 
for Milky Way stars\footnote{
The alpha element to iron abundance ratio drops 
at the ion abundance [Fe/H] $\gtrsim -1$, which is
explained by Type Ia SN chemical enrichment,
and [Fe/H] $\gtrsim -1$ corresponds to the oxygen abundance 
[O/H] $\gtrsim -0.5$ 
(i.e., $Z \gtrsim 0.3$ \Zsun)
on average for the Milky Way stars 
\citep{Nomoto2013}. 
With the iron abundance ratio scatters around the average,
the chemical enrichment of Type Ia SNe is thought to be 
negligible at [O/H] $<-1$ (i.e., $Z <0.01$ \Zsun).
}
\citep[e.g.,][]{Nomoto2013}.

Various studies actively investigate local EMPGs 
including
SBS0335-052 \citep{Izotov2009}, 
AGC198691 \citep{Hirschauer2016}, 
Little Cub \citep{Hsyu2017}, 
DDO68 \citep{Pustilnik2005}, 
I Zw18 \citep{Izotov1998}, 
Leo P \citep{Skillman2013},
and J0811+4730 \citet{Izotov2018b}.
Recently, a project named Extremely Metal-Poor Representatives Explored by the Subaru Survey \citepalias[EMPRESS;][]{Kojima2020}
has systematically studied EMPGs and examined their detailed properties.
EMPRESS uses the deep and wide multi-wavelength imaging data of the Subaru/Hyper Suprime-Cam (HSC) Survey \citep{Aihara2018}
as well as 
the Sloan Digital Sky Survey \citep[SDSS;][]{York2000}.
EMPGs are examined from a variety of perspectives,
including
high Fe/O ratios suggestive of massive stars 
(\citealt{Kojima2021}, referred to as Paper II; \citealt{Isobe2022a}, hereafter Paper IV),
morphology 
\citep[hereafter Paper III]{Isobe2021},
low-Z ends of metallicity diagnostics 
\citep[hereafter Paper V]{Nakajima2022},
outflows 
\citep[hereafter Paper VI]{Xu2022},
the shape of incident spectrum that reproduces high-ionization lines
\citep[referred to as Paper VII]{Umeda2022}, 
the primordial He abundance 
\citep[referred to as Paper VIII]{Matsumoto2022}, 
the \Ha\ kinematics 
\citep[hereafter Paper IX]{Isobe2022b},
and the resolved mass-metallicity relation (Nakajima et al. in preparation).

At $z < 0.03$, actively star-forming galaxies present 
broadband excesses in $g$- and $r$-bands that include strong nebular emission lines of \OIII$\lambda\lambda$5007,4959 and \Ha, respectively. 
In particular, 
{ an extremely}
metal-poor galaxy tends to present a red rest-frame ($g - r$) color due to a flux contribution of \OIII\ weaker than \Ha. 
Although a similar approach using a broadband excess is successfully developed for selecting intense \OIII\ emitting galaxies such as green pea galaxies at $z \sim 0.3$ \citep{Cardamone2009} and blueberry galaxies at $z\sim0.02$ \citep{Yang2017}, 
such \OIII\ emitting galaxies are not extremely metal poor because of the strong metal line (i.e., \OIII).
EMPRESS adopts a machine learning technique 
to obtain EMPG photometric candidates
showing the broadband excesses.
A similar idea is also adopted by \citet{Senchyna2019} to search for EMPGs based on the photometric data. 
Conducting follow-up spectroscopy for 
the EMPG photometric candidates,
EMPRESS has identified 11 new EMPGs with low stellar masses of $10^{4.2}-10^{6.6}$ \Msun\ 
\citepalias{Kojima2020,Isobe2022a,Nakajima2022}.

Despite decades of intensive observations, no EMPGs with metallicities below $\lesssim 0.01\ Z_\odot$ have been found so far, which is commonly referred to as the ``metallicity floor" (e.g., \citealt{Wise2012}; \citetalias{Isobe2022b}). 
We refer galaxies with Z $< 0.01$ \Zsun\ to as hyper metal poor galaxies (HMPGs).
This metallicity floor has also been observed in the stellar mass-metallicity relation of Local Group dwarf galaxies \citep[e.g.,][]{Kirby2013}. However, cosmological hydrodynamic simulations of isolated dwarf galaxies and satellites of the Milky Way analogs predict much lower stellar metallicities than what is observed \citep[e.g.,][see also \citealt{Hayashi2022}]{Wheeler2019,Applebaum2021}. It is still unknown whether HMPGs really do not exist in the local universe, or if HMPGs are too faint and/or rare to be detected in existing surveys.

Local EMPGs are often regarded as local analogs of high-$z$ primordial galaxies \citepalias[e.g.,][]{Isobe2022b} because local EMPGs have features similar to those of the simulated primordial galaxy at $z \gtrsim 7$ \citep{Wise2012}, with respect to low metallicities of 0.02--0.1 \Zsun, low stellar masses of $\sim 10^{4}$ -- $10^{8}$ \Msun, and high specific star formation rates (sSFRs) of $\sim 1$ -- 400 Gyr$^{-1}$ \citepalias{Nakajima2022}. 
Although some properties, such as size-mass relation, are different from those of high-$z$ galaxies \citepalias[e.g.,][]{Isobe2021},
local EMPGs are widely thought to be useful for understanding how young galaxies form and evolve in the early universe, 
as living fossils in the universe today. 
Furthermore, the technique we have developed to search for EMPGs in the local universe can be extended toward high-$z$ universe using the deep infrared multi-wavelength imaging data now available thanks to JWST/NIRCam
\citep[e.g.,][see also \citealt{Trussler2022}]{Naidu2022,Donnan2023,Harikane2022c}. 
We can directly compare the properties of EMPGs at different redshifts and clarify the similarities and differences. Such high-$z$ work will also allow us to examine a presence of the metallicity floor even at high-$z$ universe and its physical origin in the local universe. Accordingly, a better understanding of the properties as well as the formation process of local EMPGs is crucial to be compared with the high-$z$ counterparts and
understand galaxies in the early phase of star formation.

Physical origins of local EMPGs are discussed in various studies, and two scenarios are suggested.
One is that EMPGs have episodic star-formation histories where metal-poor gas falling on metal-enriched galaxies trigger metal-poor star formation that appear to be EMPGs, which we refer to as an ``episodic star-formation scenario"
\citep[e.g.,][]{SA14,SA15,SA16,McQuinn2020}.
The other is that EMPGs are experiencing the onset of initial star formation with extremely metal-poor gas similar to the primordial gas,
which we refer to as a ``first star-formation scenario"
\citep[e.g.,][]{Recchi2001,Isobe2022a}.

On the basis of morphologies and metal's spatial distributions,
the episodic star-formation scenario is supported
in the majority of EMPGs.
Many EMPGs are composed of
star-forming metal-poor clumps and long diffuse structures 
(e.g., \citealt{SA15}; \citetalias{Isobe2021}),
the latter of which are 
referred to as ``tails" \citepalias{Isobe2021}\footnote{
\citetalias{Isobe2021} shows that 23 out of 27 EMPGs have tails.}.
\citet{SA15} find spatial inhomogeneities of metals in 9 out of 10 EMPGs, 
where tails have metals richer than the star-forming metal-poor clumps by $\sim$ 1 dex.
\citet{SA15} conclude that metal-poor gas accretion onto the metal-enriched galaxies 
produce EMPGs with star-forming metal-poor clumps and metal-rich tails (i.e., metal-enriched galaxies).

From the point of view of chemical abundances,
the physical origins of EMPGs are still controversial.
\citet{SA16} (hereafter SA16)
select 196 EMPGs with strong emission lines, and show that about 10\% of the 196 objects have high N/O and low O/H ratios. 
These high N/O ratios can be explained by 
old gas populations that are already affected by Type Ia SNe and AGB stars, supporting
the episodic star-formation scenario.
On the other hand, 
\citetalias{Isobe2022a} points out that 
two EMPGs, J0811$+$4730 and J1631$+$4426, have high Fe/O and low N/O ratios 
that cannot be explained by the episodic star formation with chemical enrichment of 
Type Ia SNe requiring a few 100 Myr delay time
but by the onset of initial star formation 
whose gas mainly enriched by bright hypernovae (BrHNe) or pair instability supernovae (PISNe).
The most of the remaining EMPGs investigated in \citetalias{Isobe2022a} have low Fe/O and N/O ratios, which can be explained by either the episodic or first star-formation scenarios.

Investigating environments of EMPGs is 
also important 
to understand physical origins of EMPGs.
Extremely metal-poor gas may exist in low-density regions in the universe,
where intergalactic medium (IGM) is less susceptible to chemical enrichment by star formation.
In fact, several studies indicate that EMPGs reside in low-density regions 
(e.g., \citealt{Filho2015,SA16}; \citetalias{Kojima2020}).
One might think that this support the first star-formation scenario.
However, given that EMPGs are low-mass galaxies, 
it is natural that EMPGs are found in low density environment, in the framework of the $\Lambda$ cold dark matter ($\Lambda$CDM) structure formation model, 
and we cannot distinguish between the two scenarios of the episodic and first star formation.
Angular and spatial correlation functions are useful 
to quantify distributions of EMPGs and other low-mass galaxies in the framework of the structure formation model,
breaking the degeneracy between the two scenarios 
via dark-matter halo properties.
However, no correlation functions for EMPGs are derived so far.

In this paper, 
we search for metal-poor galaxy candidates 
from SDSS imaging data
using a machine-learning technique
and conducting follow-up spectroscopy.
We discuss the physical origin of EMPGs 
through the statistical analysis of clustering and chemical properties
based on a large sample of spectroscopically-confirmed EMPGs in the local universe.
We also extend our technique to a high-$z$ EMPG search using the recently available deep JWST/NIRCam imaging data to search for metal-poor galaxy candidates at $z\sim$ 4--5, 
and quickly report and discuss their properties such as metallicities.
This paper is the eleventh paper of EMPRESS.

This paper is organized as follows.
We describe the details of our selection method for identifying local EMPGs in Section \ref{sec:searching} and present the results of spectroscopic follow-up observations used to measure the metallicities of EMPG candidates. In Section \ref{sec:sample}, we present our sample of spectroscopically-confirmed EMPGs that we use in the subsequent analysis. In Section \ref{sec:clustering}, we investigate the clustering properties of our EMPG sample, while in Section \ref{sec:MZR}, we examine their chemical properties. We extend our technique to search for high-$z$ EMPGs in Section \ref{sec:jwst}. In Section \ref{sec:discussion}, we discuss the implications of our results on the existence of the metallicity floor and the physical origin of EMPGs. 
Finally, we summarize our findings and discussions in Section \ref{sec:summary}. 
Throughout the paper, we assume a solar metallicity \Zsun\ as \Oabundance\ $=$ 8.69 \citep{Asplund2021}, and adopt a standard $\Lambda$CDM cosmology with $\Omega_\Lambda = 0.7, \Omega_M = 0.3$ and $H_0 = 70$ km s$^{-1}$ Mpc$^{-1}$. All magnitudes are given in the AB system \citep{Oke1983}.

\section{Searching for EMPGs}\label{sec:searching}
This section presents local EMPGs identified by our selection method and spectroscopic observations.
We develop our selection method using a machine learning technique,
testing the selection method with the SDSS spectroscopic objects (Section \ref{sec:ml-test}).
We then apply the selection method to the faint SDSS photometric objects with no spectroscopic identifications to make EMPG photometric candidates (Section \ref{sec:photo-selection}). 
Because stars with no emission lines contaminate the EMPG photometric candidates,
we conduct shallow spectroscopic screening observations for the EMPG photometric candidates via strong emission lines with 2--4m telescopes of Seimei and Nayuta (Sections \ref{sec:seimei-screening} and \ref{sec:nayuta-screening}). 
We perform deep spectroscopy with a large telescope of 6.5m Magellan for the screened candidates to confirm EMPGs via metallicity measurements with faint emission lines detected with Magellan (Section \ref{sec:mage-obs}).

\subsection{Selection Method}
\subsubsection{Developing Machine Learning Classifier}\label{sec:ml_catalog}
In this study, we focus on identifying EMPGs at redshifts $z<0.3$ 
with a large rest-frame equivalent width of \Ha, denoted as EW$_0$(H$\alpha$), of $\gtrsim 800$ \AA. Our motivation for this selection criterion is to identify local counterparts of high-$z$ low-mass galaxies with high sSFRs of $\gtrsim 10\ \mathrm{Gyr}^{-1}$ \citep[e.g.,][]{Ono2010,Harikane2018,Stark2017}. Galaxies with such large EW$_0$(H$\alpha$) are thought to have undergone a very early phase of galaxy evolution \citep[e.g.,][]{Inoue2011}.

To identify these EMPGs efficiently from objects in a photometric catalog, we construct a machine-learning classifier using a supervised machine learning algorithm. We utilize LightGBM \citep{Ke2017}, a gradient boosting framework that employs a decision-tree based learning algorithm. The inputs for the classifier are four colors of SDSS $ugriz$ bands ($u-g, g-r, r-i$, and $i-z$) for each object, while the output is one of the four object types: EMPG, non-EMPG galaxy, star, or QSO. 
Our goal is to isolate EMPGs from other object types with minimal contamination. We chose to use colors as inputs to capture information about the spectral energy distribution of each object and avoid biases that may arise from other observable properties like size, morphology, or brightness.

\begin{figure*}
    \epsscale{1}
    \plotone{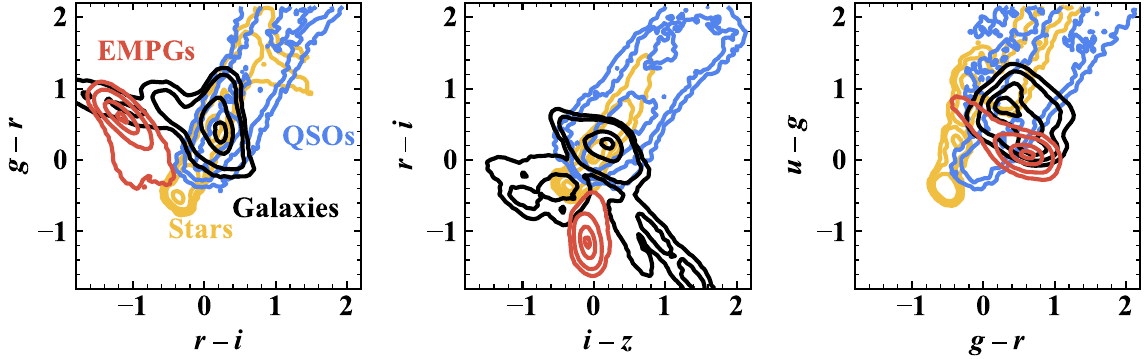}
    \caption{
     Colors of the mock observational objects
     on the color-color diagrams of 
     $g-r$ vs. $r-i$ (left), $r-i$ vs. $i-z$ (middle), and $u-g$ vs. $g-r$ (right).
     The contours show the 1, 10, 50, and 90\% level
     for each object type of
     EMPGs (red), non-EMPG galaxies (black), stars (yellow), and QSOs (blue).
    \label{fig:ccd}}
\end{figure*}

To develop the machine-learning classifier, a training sample is prepared with spectral energy distribution (SED) models, following the approach of \citetalias{Kojima2020}. The training sample consists of colors obtained from SED models of EMPGs, non-EMPG galaxies, stars, and QSOs, which are calculated by convolving the SEDs with the throughput curves of the SDSS broadband filters \citep{Fukugita1996}. The EMPG SED models used in this study are a combination of the EMPG models of \citetalias{Kojima2020} and \citet{NM22}, whose metallicity ranges are $Z =$  0.01--0.1 \Zsun\ and $Z =$ 0.001--0.01 \Zsun, respectively. The EMPG SED models
for metallicities $Z =$ 0.01--0.1 \Zsun are produced with the SED model code {\sc beagle} \citep{Chevallard2016}.
The {\sc beagle} code calculates the stellar continuum and nebular emission 
using \citet{Bruzual2003}'s stellar population synthesis code 
and the nebular emission library of \citet{Gutkin2016}, 
which is computed using the photoionization code {\sc Cloudy} \citep{Ferland2013}.
A \citet{Chabrier2003} stellar initial mass function (IMF) is applied in the {\sc beagle} code. 
We use five parameters: 
stellar mass $M_*$, age $t_\mathrm{max}$, metallicity $Z$, ionization parameter $U$, and redshift $z$.
These parameters are varied in the following ranges: $\log(M_/M_\odot) = 4.0-9.0$, $\log(t_\mathrm{max}/\mathrm{yr}) =$ 6.00--7.75, $Z=$ 0.01 -- 0.1 \Zsun, $\log U=$ $-2.7$ -- $-2.3$, and $z=$ 0.01--0.02, respectively. These ranges of parameters are based on typical values observed in EMPGs and/or theoretical predictions.
The EMPG SED models for metallicities $Z =$ 0.001--0.01 \Zsun\ are based on the SED models presented by \citet{NM22}. These models have metallicities of $Z = [1/1400, 1/140]$ \Zsun, cover stellar ages from 6.0 $<$ log(age/yr) $<$ 7.8 with a step size of 0.2, span ionization parameters $\log U$ from -3.0 to -1.0 with a step size of 0.5, and include dust attenuations of $E(B-V) =$ [0, 0.01, 0.02, 0.05], assuming a \citet{Kroupa2001} IMF with upper mass limits of 100 \Msun\ and 300 \Msun.
The SED models of non-EMPG galaxies are generated 
similarly
to those of EMPGs { with $Z =$ 0.01--0.1 \Zsun}, but with different parameter ranges,
which contains the SED models of high-$z$ ($z=$ 0.08--0.40) \OIII\ emitters.
The stellar SED models are based on those of \citet{Castelli2004}.
The SED model of QSOs are produced
from a observed composite spectrum of QSOs at $1 < z < 2.1$ \citep{Selsing2016}
by varying three parameters:
power-law index ($\alpha$) of an intrinsic near-ultraviolet slope $f_\lambda\propto\lambda^\alpha$, 
$V$-band dust attenuation optical depth, 
and redshift.
Full details of the SED models are described in \citetalias{Kojima2020}.
The numbers of the model SEDs are 4,752, 6,728, 4,240, and 7,250 for the EMPGs, non-EMPG galaxies, stars, and QSOs, respectively. 
We add random noise to the colors to simulate observational errors, and mock observational objects are produced. The mock EMPGs, non-EMPG galaxies, stars, and QSOs consist of 475,200, 672,800, 42,400, and 72,500 objects, respectively. From each of the four object types, 30,000 mock observational objects are randomly selected to create a training sample of 120,000 objects in total ($=30,000 \times 4$).
Figure \ref{fig:ccd} shows the colors of the mock observational objects
on the color-color diagrams of $g-r$ versus $r-i$, $i-z$ versus $r-i$, and $g-r$ versus $u-g$.

We divide the training sample into two sets: 80\% for training and the remaining 20\% for validation. We train the machine-learning classifier to minimize classification errors on the training data set. 
LightGBM has several hyperparameters that control the learning process. We focuse on five hyperparameters:
{\tt num\_leaves}, {\tt min\_data\_in\_leaf}, {\tt learning\_rate}, {\tt bagging\_fraction}, {\tt bagging\_freq}, 
which are important in improving classification performance while avoiding over-fitting to the training data set using an automatic hyperparameter optimization software framework, optuna\footnote{\url{https://optuna.org/}}.
The meanings of these hyperparameters can be found in the LightGBM documentation\footnote{\url{https://lightgbm.readthedocs.io/en/v3.3.4/pythonapi/lightgbm.LGBMClassifier.html}}.
Default values are used for the remaining hyperparameters.
To avoid bias towards a particular model, we select 20 hyperparameter sets that show comparable performance. We then design our classifier to select EMPG candidates predicted by at least a certain number of models. We set this threshold at 3 because it yields the highest number of true EMPGs selected in the validation data set.
To evaluate the performance of the classifier, we calculate both accuracy and F1 score on the validation data set, which both metrics yielding results over 0.99. Here, the accuracy is determined by the total number of correctly identified EMPGs (true positives) as well as the correctly identified non-EMPG galaxies, stars, or QSOs (true negatives), which are then divided by the total number of classifications. This can be expressed as ($\mathrm{N_{TP}}$+$\mathrm{N_{TN}}$)/$\mathrm{N_{tot}}$, where $\mathrm{N_{TP}}$ ($\mathrm{N_{TN}}$) represents the number of actual EMPGs (non-EMPG galaxies, stars, or QSOs) that are correctly classified as such. The F1 score is also a widely used metric for evaluating binary classification tasks, and it is calculated as the harmonic mean of precision and recall. Precision measures the proportion of true positives out of the total number of samples that are classified as EMPGs, while recall measures the proportion of true positives out of the total number of actual EMPGs.

\subsubsection{Testing the Classifier with the SDSS Spectroscopic Data} \label{sec:ml-test}
Before we apply the machine learning classifier to the SDSS photometric data, 
we test whether the classifier correctly isolates EMPGs from other-type objects (non-EMPG galaxies, stars, or QSOs)
with the SDSS objects having spectroscopic classifications (i.e., galaxy, star, or QSO) and emission-line measurements, as well as optical photometry.
Using the SDSS DR16 \citep{Ahumada2020} data,
we select objects whose photometric measurements are brighter than the SDSS limiting magnitudes (95\% completeness for point sources), $u < 22.0$ mag, $g< 22.2$ mag, $r<22.2$ mag, $i<21.3$ mag, and $z<21.3$ mag. 
We remove objects with errors larger than $0.2$ mag in $u$, $g$, $r$, $i$, and $z$ bands. 
Note that we use {\tt Modelmag} in the SDSS DR16 data.
We remove an object with ``0" (i.e., unclean) for the clean flag in {\tt PhotoObjALL} of the SDSS DR16 data that has a photometric measurement problem, such as a duplication, deblending/interpolation, suspicious detection, and detection at the edge of an image.
We also remove an object with $r - i > 0$ whose color is apparently different from EMPGs (left and middle panels of Figure \ref{fig:ccd})
to save computational power.
By these selections, we have constructed a catalog 
with the spectroscopy and photometry
composed of 24,419 sources in total
(998 galaxies, 22,583 stars, and 838 QSOs) 
that is referred to as the SDSS test catalog.

We apply our machine-learning classifier to the SDSS test catalog, and obtain 40 EMPG candidates.
We check the spectroscopic classifications (i.e., galaxy, star, and QSO) of these EMPG candidates,
and identify 38 candidates as galaxies and 2 candidates as stars.
Of the 38 galaxies, we find detailed spectroscopic studies for 18 galaxies in the literature
(\citealt{Kniazev2003,Izotov2006b,Izotov2012a,Thuan2005,Guseva2007}; \citetalias{Nakajima2022}).
We obtain direct-method metallicity measurements based on \OIII$\lambda$4363 of these 18 galaxies.
We find that 6 out of 18 galaxies are EMPGs (\Oabundance\ $=$ 7.22--7.65).
The other 12 galaxies are also actively star-forming galaxies with intense emission lines 
but have slightly higher metallicities (\Oabundance\ $=$ 7.72--8.37).
Hereafter, we refer to such a moderately metal-poor galaxy as a ``MPG".
For the remaining 20 galaxies, we estimate metallicities with strong line ratios as described below.
We obtain 
N2 (\NII$\lambda$6584/\Ha), 
R3 (\OIII$\lambda$5007/\Hb),
and R23 ((\OIII$\lambda\lambda$5007, 4959 $+$ \OII$\lambda$3727)/\Hb)
indices using the flux measurements in the SDSS {\tt galSpecLine} catalog, only for sources with S/N $>$ 3 for all emission lines used. 
Using the metallicity calibrations of \citetalias{Nakajima2022}, 
we determine the high- or low- metallicity branch (i.e., whether \Oabundance\ $\gtrsim$ 8 or not) from N2 index, 
and then we estimate metallicities from R23 index if available, otherwise R3 index.
We find that 5 are EMPGs (\Oabundance\ $\simeq$  7.54--7.60),
13 are MPGs (\Oabundance\ $\simeq$  7.69 -- 8.61),
and the remaining one is metal-rich galaxies (\Oabundance\ $\simeq$ 8.80).
Taken together, the { purity} is 28\% (11/40),
which is lower than the accuracy in the training (Section \ref{sec:ml_catalog})
but comparable to the results of spectroscopic observations in \citetalias{Kojima2020}, 30\% (3/10).

\subsubsection{Comparison between Classifiers}
In this section, we present a comparison of the selection performance of our machine-learning classifier with those of \citetalias{Kojima2020} and \citetalias{Nakajima2022}.
These three methods share the same idea; they isolate EMPGs based on photometric data using color excess expected for EMPGs at $z < 0.03$ with $EW_0(H\alpha) > 800$ \AA.
One difference is the choice of a machine learning algorithm.
While \citetalias{Kojima2020} and \citetalias{Nakajima2022} use CNN, we use lightGBM,
which is widely used for classification tasks on tabular data.
{We can} see how much the performance would change with a classifier using a different algorithm other than CNN.
Another difference is the training data.
\citetalias{Nakajima2022} and we use the SED models 
of galaxies with $Z < 0.01$ \Zsun\ { (i.e., HMPGs)} in the training data, while \citetalias{Kojima2020} does not, which allow us to search for { HMPGs}.

We investigate whether these three classifiers produce differences in a selection performance.
We select EMPG candidates using the classifier of \citetalias{Kojima2020} (\citetalias{Nakajima2022}) from the SDSS test catalog described in Section \ref{sec:ml-test}. A total of  47 (40) objects are selected as EMPG candidates, of which 42 (38) are galaxies and 5 (2) are stars based on the SDSS spectroscopic classifications. We obtain direct-method metallicities for 20 (18) out of the 42 (38) galaxies from the literature
(\citealt{Kniazev2003,Izotov2006b,Izotov2012a,Thuan2005,Guseva2007}; \citetalias{Nakajima2022}), and find that 6 (7) are EMPGs and 14 (11) are MPGs. For the remaining 22 (20) objects, we estimate metallicities using the strong line method as described in Section \ref{sec:ml-test}, resulting in 4 (4) being EMPGs, 15 (13) being MPGs, and { 3 (3) being metal-rich galaxies}. The overall { purity} of the classifiers of \citetalias{Kojima2020} and \citetalias{Nakajima2022} are 21\% (10/47) and 28\% (11/40), respectively. The results are summarized in Table \ref{tab:ml-test}.

Compared to \citetalias{Kojima2020}, our and \citetalias{Nakajima2022}'s classifiers reduce the rate of stellar contamination from 10\% to 5\%, while maintaining a similar number of correctly selected EMPGs. Although there is some overlap in the EMPGs selected by each classifier, we select 3 (2) EMPGs that are not selected by the classifier of \citetalias{Kojima2020} (\citetalias{Nakajima2022}), while the classifier of \citetalias{Kojima2020} (\citetalias{Nakajima2022}) select 2 (2) EMPGs that are not selected by ours. Combining multiple models may offer the advantage of increasing the total number of correctly selected EMPGs.

\begin{table}
\caption{The number of EMPGs, MPGs, and stars that are selected by the each mahine-learning classifier from the SDSS test catalog.}
\label{tab:ml-test}
\begin{center}
\begin{tabular}{cccc}
\hline\hline
& This study & Paper I & Paper V \\ \hline
EMPG & 11 (6) & { 10 (6)} & { 11 (7)} \\
{ Other galaxies} & 27 (12) & { 32 (14)} & { 27 (11)} \\
Star & 2 & { 5} & { 2} \\ \hline
Total & 40 & { 47} & { 40} \\ \hline
\end{tabular}
\end{center}
\footnotesize{{\sc Note}---The number of EMPGs and MPGs with the direct-method metallicity measurements are noted in brackets.}
\end{table}

\subsection{Photometric Selection} \label{sec:photo-selection}
In this section, we select EMPG candidates from an SDSS photometric catalog with our machine-learning classifier, which works well with the SDSS test catalog (Section \ref{sec:ml-test}). 
Based on the SDSS DR16 \citep{Ahumada2020} photometric data 
with no SDSS spectroscopic measurements,
we select objects
that satisfy the criteria of
the SDSS limiting magnitudes, magnitude errors, flags, and colors
as described in Section \ref{sec:ml-test}.
We obtain 569,825 sources in total from the SDSS DR16 photometric data that is referred to as the SDSS photometric catalog.
We apply our machine-learning classifier to the SDSS photometric catalog and 734 objects are classified as EMPGs.
We check the 
images of the 734 candidates 
and eliminate 
sources that obviously have an unclean photometric data 
but not fully removed with the flags.
By this visual inspection, 
we obtain 134 EMPG candidates, whose $i$-band magnitudes are in the range of $i \simeq$ 14 -- 20 mag. 

\subsection{Spectroscopic Screening} \label{sec:spec-screening}
We carried out spectroscopic screening observations with 3.8m Seimei and 2m Nayuta telescopes
for the 134 candidates
to remove contamination of stars via strong emission lines
and select candidates for deep spectroscopy.

\subsubsection{Screening with 3.8m Seimei} \label{sec:seimei-screening}
We carried out spectroscopic observations for 130 EMPG candidates 
with KOOLS-IFU \citep{Matsubayashi2019} on the Seimei telescope \citep{Kurita2020}
during 2020 December and 2021 March (PI: Y. Isobe). 
We used the VPH-blue grism 
with a wavelength range of 4100--8900 \AA\
and a spectral resolution power of $R\sim500$.
The exposure times were 600 s.
The sky was clear during the observations 
with seeing sizes of around $2''$.

\begin{figure}
    \epsscale{1}
    \plotone{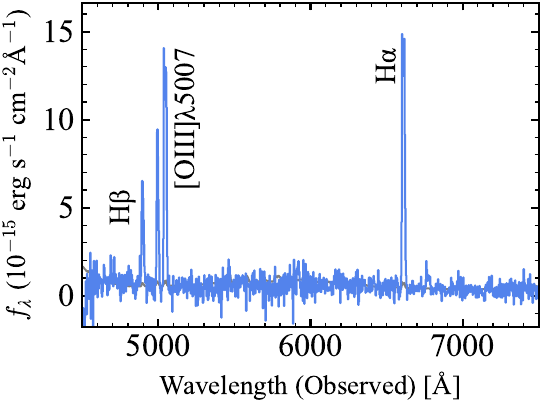}
    \caption{Spectrum of J1034--0221 obtained by Seimei observations. The gray line indicates the error spectrum.
    \label{fig:seimei}}
\end{figure}

Data reduction is conducted using the software 
provided on the website of KOOLS-IFU\footnote{\url{http://www.kusastro.kyoto-u.ac.jp/~kazuya/p-kools/reduction-201806/index.html}}.
The reduction and calibration processes include the bias subtraction, flat-fielding, sky subtraction, wavelength calibration, and flux calibration. 
A 1D spectrum is made by combining spectra of the region where a target flux is detected.
A 1D error spectrum contains read-out and photon noise of sky and object emission.
One of the reduced spectra obtained by the Seimei observations is shown in Figure \ref{fig:seimei}.
We note that the flux of the \OIII$\lambda$5007 line in Figure \ref{fig:seimei} may be saturated. However, the flux value is not important since the purpose of the screening observations is to isolate galaxies with strong emission lines.

\subsubsection{Screening with 2m Nayuta} \label{sec:nayuta-screening}
\begin{figure}
    \epsscale{1}
    \plotone{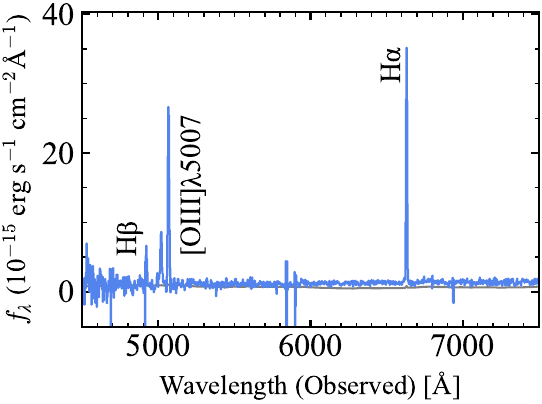}
    \caption{Spectrum of J1804$+$0008 obtained by Nayuta observations.
    \label{fig:nayuta}}
\end{figure}

We carried out spectroscopic observations for four EMPG candidates 
with MALLS on the Nayuta telescope on 2021 April (PI: Y. Isobe). 
We used the 150 mm$^{-1}$ grating with a slit width of 1.2 arcsec.
Our observations had a wavelength coverage of $3700-9500$ \AA\
and a spectral resolution of $R\sim600$.
The exposure time was 1200 s. 
The seeing size during the observation was around $4''$.
Data reduction is conducted in a standard manner with {\sc IRAF}.
One of the reduced spectra obtained by the Nayuta observations is shown in Figure \ref{fig:nayuta}.

\subsubsection{Results of the Spectroscopic Screening Observations} \label{sec:results-screening}
Of the 134 sources, 104 (77) objects have a strong emission line
around wavelength of \Ha\ (\Hb\ and \OIII$\lambda$5007). 
Some sources display strong \Hb, \OIII$\lambda$5007, and \Ha\ emission lines but lack an apparent detection of \NII$\lambda6583$, 
 indicating a low \NII$\lambda6583$/\Ha\ ratio (log(\NII$\lambda6583$/\Ha) $\lesssim -1$) and corresponding to low metallicities ($Z \lesssim 0.5$ \Zsun; \citetalias{Nakajima2022}).
We select 10 sources that are observable with the Magellan telescope during our scheduled observation dates and do not have significant \NII$\lambda6583$ detections indicating low metallicities.
Out of these 10 sources, 
eight were identified during the Seimei observations, and the remaining two were identified during the Nayuta observations.

\subsection{Spectroscopic Confirmation of Metallicity} \label{sec:mage-obs}
\subsubsection{MagE Observation and Data Reduction}

\begin{table}
\caption{Summary of MagE Observations.}
\label{tab:mage}
\begin{center}
\begin{tabular}{ccccc}
\hline\hline
ID & R.A. & Dec. & Slit & Exposure \\
& (hh:mm:ss) & (dd:mm:ss) & (arcsec) & (s) \\
(1) & (2) & (3) & (4) & (5) \\ \hline
J1034--0221 & 10:34:01.0 & --02:21:50.4 & 1.0 & 600 \\
J1244+2828 & 12:44:41.3 & +28:28:04.4 & 1.0 & 600 \\
J1305+2852 & 13:05:46.9 & +28:52:03.0 & 1.0 & 600 \\
J1432+0611 & 14:32:07.3 & +06:11:16.2 & 0.70 & 300 \\ 
J1526+1610 & 15:26:11.6 & +16:10:00.8 & 1.0 & 600 \\
J1604+1459 & 16:04:18.0 & +14:59:37.5 & 1.0 & 600 \\
J1616+1453 & 16:16:29.8 & +14:53:24.1 & 1.0 & 600 \\
J1637+1729 & 16:37:42.0 & +17:29:50.7 & 0.70 & 300 \\
J1804+0008 & 18:04:19.6 & +00:08:05.8 & 0.70 & 300 \\
J2136+0414 & 21:36:58.0 & +04:14:00.0 & 1.0 & 300 \\ \hline
\end{tabular}
\end{center}
\footnotesize{{\sc Note}---(1) ID. (2) Right ascension in J2000. (3) Declination in J2000. (4) Slit width. (5) Exposure time.}
\end{table}

\begin{figure*}
    \epsscale{1}
    \plotone{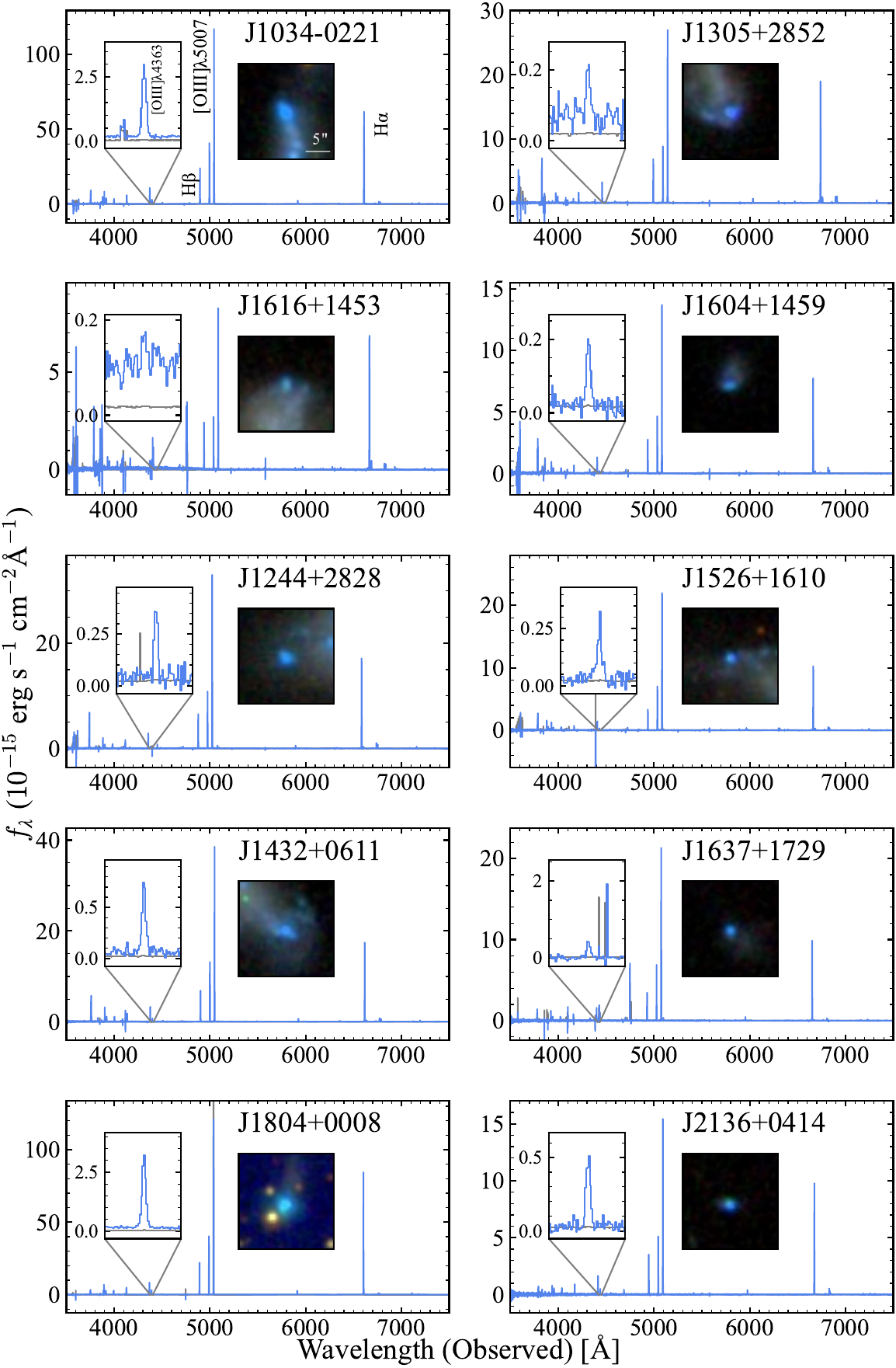}
    \caption{MagE spectra of the ten EMPG candidates. The gray lines indicate error spectra. The left-hand inset of each panel presents an enlarged view of the spectrum around \OIII$\lambda$4363. 
    The right-hand inset of each panel shows $20'' \times 20''$ cutout $gri$-composite images from SDSS.
    \label{fig:mage}}
\end{figure*}

We carried out deep spectroscopy
to confirm EMPGs via metallicity measurements with faint emission lines detected with the Magellan Echellette Spectrograph \citep[MagE;][]{Marshall2008} on the Magellan telescope.
We observed the 10 EMPG candidates from the results of screening observations (Section \ref{sec:results-screening}).
The observing nights were 2021 July 10, October { 9}, 2022 April 27, and 28 (PI: M. Rauch). 
We used the echellette grating with the $0\farcs70\times10''$ or $1\farcs0\times10''$ slits. 
The exposure times were 300 s or 600 s, depending on luminosities of targets. 
The MagE spectroscopy covered $\lambda \sim 3100-10000$ \AA\ with a spectral resolution of $R \sim 4000$.
We observed the standard stars CD329927 and Feige56.
All the nights were clear with a typical seeing sizes of $\sim 0\farcs7$.

We reduce the MagE data 
using PypeIt\footnote{\url{https://github.com/pypeit/PypeIt}}
\citep{Prochaska2020b},
an open-source -based spectroscopic data reduction package.
PypeIt performs
bias subtraction, flat fielding, sky subtraction, wavelength calibration, cosmic-ray removal, and extracting 1D spectra.
Following the procedure described in \citetalias{Xu2022},
we modify a sky model around bright extended emission lines (e.g., \Ha\ and \OIII$\lambda$5007), where PypeIt fails to fit the sky model.
Specifically, we adopt flat sky background in pixels where bright emission lines exist by averaging sky background in the nearby pixels. 
Note that most emission lines we investigate do not overlap with strong skylines.
We show MagE spectra as well as SDSS images of the 10 candidates in Figure \ref{fig:mage}. 

We measure the flux of each of the lines by fitting a Gaussian profile plus a constant continuum. 
The measured fluxes are then corrected for Galactic extinction based on the \citet{Schlafly2011}'s map as well as the extinction curve of \citet{Calzetti1989}.
We use the Balmer lines of \Ha, \Hb, \Hg, and \Hd\ to estimate the dust attenuation assuming the \citet{Calzetti2000}' attenuation curve,
following the early EMPRESS work \citepalias{Isobe2022a,Nakajima2022}.
The obtained $E(B-V)$ values are listed in Table \ref{tab:oh}. These $E(B-V)$ values are higher than those we assume in SED fitting in Section \ref{sec:searching} but comparable to those of the EMPGs identified in \citetalias{Kojima2020}.
Flux errors are estimated from 1D error spectra extracted by PypeIt.
Table \ref{tab:flux} lists the key line fluxes normalized by \Hb\ and their $1\sigma$ errors for the 10 candidates.

\begin{figure}
    \epsscale{1}
    \plotone{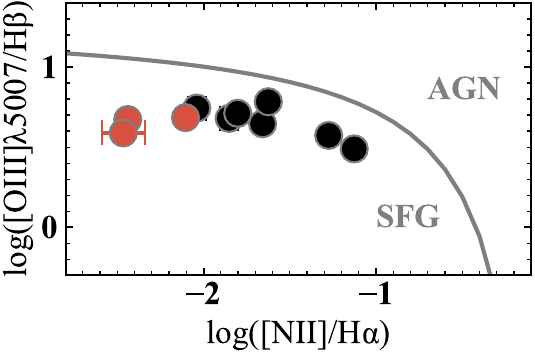}
    \caption{Ten MagE objects on the BPT diagram. The flux ratios are derived from extinction-corrected line flux of MagE spectra. The red (black) points indicate objects with metallicities below (above) 0.1\Zsun.
    The gray line present the demarcation line defined by \citet{Kauffmann2003} which divides regions dominated by star-forming galaxies (SFGs) and AGNs.
    \label{fig:bpt}}
\end{figure}

We confirm that the 10 candidates do not show an obvious active galactic nucleus (AGN) activity based on the BPT \citep{Baldwin1981} diagram \citep{Kauffmann2003}, as presented in Figure \ref{fig:bpt}. 
We note that we can not exclude the possibility of the existence of a metal-poor AGN with the BPT diagram \citep[e.g.,][]{Kewley2013}

\begin{table*}
\caption{Physical properties of the 10 EMPG candidates.}
\label{tab:oh}
\begin{center}
\begin{tabular}{clcccc}
\hline\hline
ID & Redshift & $E(B-V)$ & $T_\mathrm{e}$(\OIII) & $n_\mathrm{e}$(\OII) & \Oabundance\ \\
& & (mag) & ($10^4$K) & ($\mathrm{cm^{-3}}$) & \\
(1) & (2) & (3) & (4) & (5) & (6) \\
\hline
J1034--0221 & 0.00718 & $0.254 \pm 0.005$ & $1.67 \pm 0.01$ & $55 \pm 22$ & $7.668 \pm 0.003$ \\
J1244+2828 & 0.00337 & $0.221 \pm 0.007$ & $1.13 \pm 0.01$ & $17 \pm 6$ & $8.201 \pm 0.013$ \\
J1305+2852 & 0.02698 & $0.222 \pm 0.007$ & $0.94 \pm 0.02$ & $232 \pm 6$ & $8.376 \pm 0.025$ \\
J1432+0611 & 0.00831 & $0.154 \pm 0.008$ & $1.40 \pm 0.01$ & $70 \pm 6$ & $7.945 \pm 0.010$ \\
J1526+1610 & 0.01540 & $0.327 \pm 0.012$ & $1.21 \pm 0.02$ & $78 \pm 2$ & $8.163 \pm 0.019$ \\
J1604+1459 & 0.01517 & $0.225 \pm 0.010$ & $1.33 \pm 0.02$ & $95 \pm 5$ & $7.962 \pm 0.022$ \\
J1616+1453 & 0.01588 & $0.221 \pm 0.013$ & $0.98 \pm 0.03$ & $24 \pm 2$ & $8.311 \pm 0.042$ \\
J1637+1729 & 0.01363 & $0.254 \pm 0.014$ & $1.58 \pm 0.02$ & $72 \pm 6$ & $7.765 \pm 0.015$ \\
J1804+0008 & 0.00629 & $0.507 \pm 0.007$ & $1.90 \pm 0.01$ & $144 \pm 74$ & $7.504 \pm 0.005$ \\
J2136+0414 & 0.01714 & $0.213 \pm 0.014$ & $1.96 \pm 0.50$ & $48 \pm 63$ & $7.388 \pm 0.399$ \\
\hline
\end{tabular}
\end{center}
\footnotesize{{\sc Note}---(1) ID. (2) Redshift. (3) Dust attenuation of gas. (4) Electron temperature of {\sc O}$^{2+}$. (5) Electron density of {\sc O}$^{+}$. (6) Gas-phase metallicity.}
\end{table*}

\begin{sidewaystable*}
\caption{Extinction-corrected emission-line fluxes of the 10 EMPG candidates}
\label{tab:flux}
\begin{center}
\begin{tabular}{ccccccccccc}
\hline\hline
ID & \OII$\lambda$3727 & \OII$\lambda$3729 & \Hd\ & \Hg\ & \OIII$\lambda$4363 & \Hb\ & \OIII$\lambda$4959 & \OIII$\lambda$5007 & \Ha\ & \NII$\lambda6583$ \\
(1) & (2) & (3) & (4) & (5) & (6) & (7) & (8) & (9) & (10) & (11) \\
\hline
J1034-0221 & $30.4 \pm 0.6$ & $43.2 \pm 0.6$ & $25.6 \pm 0.2$ & $46.4 \pm 0.2$ & $11.8 \pm 0.1$ & $100.0 \pm 0.3$ & $165.3 \pm 0.3$ & $483.2 \pm 0.8$ & $272.2 \pm 0.9$ & $2.1 \pm 0.0$ \\
J1244+2828 & $74.1 \pm 0.5$ & $109.2 \pm 0.4$ & $25.0 \pm 0.4$ & $44.8 \pm 0.3$ & $4.5 \pm 0.2$ & $100.0 \pm 0.5$ & $158.8 \pm 0.7$ & $478.0 \pm 0.9$ & $276.7 \pm 0.5$ & $3.9 \pm 0.2$ \\
J1305+2852 & $73.6 \pm 0.4$ & $89.8 \pm 0.2$ & $22.8 \pm 0.4$ & $45.9 \pm 0.4$ & $2.0 \pm 0.2$ & $100.0 \pm 0.5$ & $124.4 \pm 0.6$ & $374.2 \pm 1.1$ & $280.4 \pm 0.6$ & $14.9 \pm 0.1$ \\
J1432+0611 & $64.0 \pm 0.5$ & $90.0 \pm 0.3$ & $24.3 \pm 0.4$ & $45.5 \pm 0.4$ & $8.7 \pm 0.3$ & $100.0 \pm 0.6$ & $184.1 \pm 0.6$ & $514.6 \pm 1.1$ & $271.7 \pm 0.6$ & $4.3 \pm 0.1$ \\
J1526+1610 & $61.5 \pm 0.1$ & $85.7 \pm 0.2$ & $21.9 \pm 0.6$ & $44.0 \pm 0.8$ & $7.1 \pm 0.4$ & $100.0 \pm 0.8$ & $204.3 \pm 1.0$ & $607.3 \pm 1.8$ & $271.2 \pm 0.9$ & $6.4 \pm 0.2$ \\
J1604+1459 & $69.8 \pm 0.3$ & $96.1 \pm 0.4$ & $23.2 \pm 0.5$ & $43.5 \pm 0.6$ & $6.5 \pm 0.5$ & $100.0 \pm 0.7$ & $149.3 \pm 0.4$ & $442.6 \pm 0.5$ & $269.0 \pm 0.3$ & $5.9 \pm 0.1$ \\
J1616+1453 & $91.3 \pm 0.3$ & $133.8 \pm 0.1$ & $22.8 \pm 0.7$ & $46.2 \pm 0.7$ & $1.9 \pm 0.3$ & $100.0 \pm 0.9$ & $102.1 \pm 1.3$ & $308.3 \pm 1.1$ & $281.3 \pm 0.8$ & $21.0 \pm 0.3$ \\
J1637+1729 & $27.7 \pm 0.1$ & $38.9 \pm 0.2$ & $28.6 \pm 0.9$ & $47.9 \pm 1.0$ & $12.1 \pm 0.5$ & $100.0 \pm 0.9$ & $184.3 \pm 0.7$ & $554.5 \pm 1.9$ & $282.8 \pm 0.8$ & $2.6 \pm 0.1$ \\
J1804+0008 & $13.6 \pm 0.8$ & $18.2 \pm 0.7$ & $26.3 \pm 0.2$ & $45.6 \pm 0.4$ & $14.9 \pm 0.2$ & $100.0 \pm 0.5$ & $176.0 \pm 0.7$ & $470.3 \pm 1.0$ & $271.8 \pm 0.8$ & $1.0 \pm 0.0$ \\
J2136+0414 & $14.5 \pm 1.1$ & $20.7 \pm 1.3$ & $25.6 \pm 0.9$ & $44.8 \pm 0.8$ & $12.7 \pm 0.5$ & $100.0 \pm 0.9$ & $130.4 \pm 0.9$ & $387.9 \pm 1.5$ & $269.1 \pm 0.7$ & $0.9 \pm 0.1$ \\
\hline
\end{tabular}
\end{center}
\footnotesize{{\sc Note}---(1) ID. (2) -- (11) Extinction-corrected flux and $1\sigma$ error normalized by the \Hb\ flux and error.}
\end{sidewaystable*}

\subsubsection{Metallicities}
We identify the temperature-sensitive line of \OIII$\lambda$4363 in all of the 10 EMPG candidates,
which allows us to derive gas-phase metallicities with the direct $T_\mathrm{e}$ method as described below. 
We use the  tool PyNeb \citep{Luridiana2015} to estimate 
electron temperatures, electron number densities, and ion abundances.
We use \OIII$\lambda$4363/\OIII$\lambda\lambda$4959,5007 ratio and \OII$\lambda\lambda$3726/3729 ratio
to estimate electron temperatures and electron number densities of O$^{2+}$ ($T_\mathrm{e}$(\OIII) and $n_\mathrm{e}$(\OII))
using the PyNeb package {\tt getCrossTemDen}. 
The temperature of O$^+$, $T_\mathrm{e}$(\OII), is estimated from $T_\mathrm{e}$(\OIII) using an empirical relation of \citet{Garnett1992}. 
We assume that the density of the gas is uniform throughout.
We derive the abundances of {\sc O$^+$/H$^+$} 
with the \OII$\lambda\lambda$3726,3729 / \Hb\ ratio and $T_\mathrm{e}$(\OII),
and {\sc O$^{2+}$/H$^+$} with the \OIII$\lambda\lambda$4959,5007 / \Hb\ ratio and $T_\mathrm{e}$(\OIII), using the PyNeb package {\tt getIonAbundance}. 
We ignore O$^{3+}$ and higher-order oxygen ions as in previous work \citepalias[e.g.,][]{Isobe2022a}.
The measured oxygen abundances as well as redshift, $E(B-V)$, and $T_\mathrm{e}$(\OIII) are summarized in Table \ref{tab:oh}. 

The derived metallicities are in the range of $Z=$ 0.05--0.5 \Zsun\
as similarly identified in earlier studies of EMPRESS \citepalias{Kojima2020,Isobe2022a,Nakajima2022}.
Three of the ten EMPG candidates have a metallicity below the 10\% solar value (12 + log(O/H) $<$ 7.69) and are thus confirmed to be EMPGs, 
but still no object with metallicity below 0.01 \Zsun\ is found.
The remaining seven sources are turned out to 
have slightly higher metallicities (0.1--0.5 \Zsun).

\section{EMPG Sample}\label{sec:sample}
Here we describe our EMPG sample.
Including the three EMPGs newly identified in this study (Sec. \ref{sec:searching}), 
EMPRESS has delivered 14 EMPGs so far below $Z<0.1$ \Zsun 
whose stellar masses are particularly low 
\citepalias[$M_*\sim10^4$--$10^8$ \Msun;][]{Kojima2020,Isobe2021,Nakajima2022,Xu2022}.
Adding the other 89 local EMPGs that are compiled from the literature \citepalias{Nakajima2022},
we use the sample of 105 EMPGs in the following statistical analysis.
This is the largest sample constructed below $Z<0.1$ \Zsun\
fully based on the direct temperature method.
Basic key properties of the 105 EMPGs, 
such as metallicity, stellar mass ($i$-band magnitude), and SFR (UV magnitude and Balmer emission) are summarized in \citepalias{Nakajima2022}. 

\section{Clustering}\label{sec:clustering}
To investigate clustering properties of EMPGs,
we derive the projected cross-correlation function (CCF), $\omega_{\rm p}(r_{\rm p})$, of our EMPG sample.

For our cross-correlation analysis, we use the large-scale structure catalog of \citet{Reid2016}.
Although the catalog of \citet{Reid2016} is not large,
we can exploit the random catalog given by this study. 
We obtain a sample of 30,757 galaxies in the EMPG redshift range of $z=0-0.1$, 
which 
correspond to the all galaxies that lie in the redshift range of $0 < z < 0.1$ from the catalog of \citet{Reid2016}, and
are referred to as the cross-correlation analysis (CCA) galaxies.
The CCA sample is close to volume-limited over this redshift range \citep{Reid2016}.

Of our EMPG sample,
we only use 50 galaxies 
{ which are located in the same region as the CCA sample.}
Because we need a large sample to obtain a CCF with high statistical accuracy, we add 166 EMPGs identified by \citetalias{SA16} to our EMPG sample to increase the size of the sample.
The 166 EMPGs exist in the { same region as the} CCA sample.
We refer to our EMPG sample with the \citetalias{SA16} EMPGs as the extended EMPG sample.
Note that the metallicities of the \citetalias{SA16} EMPGs have not been measured by the direct method with {\sc [Oiii]}$\lambda$4363 emission, but the method with the strong lines.
\citetalias{SA16} claim that the systematic uncertainty of the strong-line method against the direct method is estimated to be 0.04 dex that is smaller than the statistical uncertainty. Thus the contamination of metal rich galaxies is negligibly small in the additional 166 EMPGs.
The extended EMPG sample contains 216 EMPGs in total.

We also construct a comparison sample from the SDSS galaxies with the stellar mass distribution same as the EMPGs. 
We use stellar masses in the SDSS {\tt galSpecExtra} catalog for the SDSS galaxies and the 166 EMPGs identified by \citetalias{SA16}, while we estimate stellar masses of the remaining 50 EMPGs as described in Section \ref{sec:MZR}.
We select galaxies with any metallicities at our EMPGs redshift range of $z=0-0.1$ from the SDSS DR12 spectroscopic catalog of \citet{Alam2015}. We then apply the criteria of stellar masses that should fall within the 68th percentile of the stellar mass distribution of the extended EMPGs sample
because some EMPGs have low stellar masses of $M_* < 10^{6}$ \Msun,
which are not covered by that of SDSS galaxies.
We obtain the comparison sample consisting of 1,881 galaxies. The stellar masses of the extended EMPGs (comparison) sample are in the range of $\log (M_*/$\Msun) $=$ 3.7--10.3 (7.0--8.4).

\begin{figure}
    \epsscale{1}
    \plotone{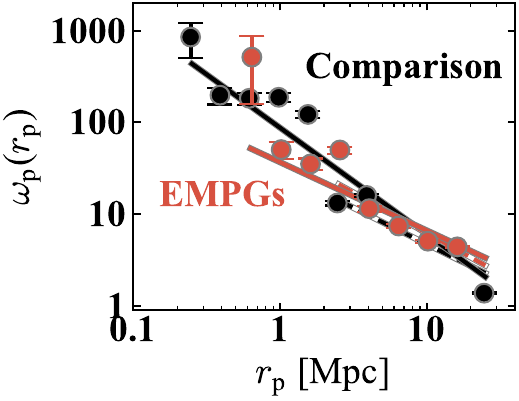}
    \caption{CCFs for the extended-EMPG and comparison samples. 
            The red (black) points denote $\omega_{\rm p}(r_{\rm p})$ of the extended-EMPG (comparison) sample.
            The red (black) solid line shows the best-fit power-law functions of the extended-EMPG (comparison) sample.
            The red (black) dashed line is the same as the red (black) solid line, but for the CCF data points only in the large scale of $r_{\rm p} > 2$ Mpc.
    \label{fig:ccf_2d}}
\end{figure}

We calculate the galaxy CCF between EMPGs and CCA galaxies on a two-dimensional grid of pair separations parallel ($\pi$) and perpendicular ($r_{\rm p}$) to the line of sight. 
We estimate $\xi(r_{\rm p}, \pi)$ using the \citet{Landy1993} estimator,
\begin{equation}
    \xi(r_{\rm p}, \pi) = \frac{D_1D_2 - D_1R_2 - R_1D_2 + R_1R_2}{R_1R_2},
\end{equation}
where $DD$, $DR$, $RD$, and $RR$ are the normalized numbers of galaxy-galaxy, galaxy-random, random-galaxy, and random-random pairs in each separation bin, 
and the subscripts denote the two subsamples.
We use the random catalog of \citet{Reid2016} for the CCA galaxies.
We then compute the projected correlation function,
\begin{equation}
    \omega_{\rm p}(r_{\rm p}) = 2\int_0^\infty d\pi\xi(r_{\rm p}, \pi).
\end{equation}
In practice we integrate up to $\pi=40\ h^{-1}$Mpc, following \citet{Zehavi2005}. 
The errors are estimated on the basis of the Poisson statistics,
\begin{equation}
    \sigma = \frac{1+\omega_{\rm p}(r_{\rm p})}{\sqrt{D_1D_2}}. \label{eq:poisson}
\end{equation}
We also obtain the CCF of the comparison sample 
in the same manner as the one of the extended EMPG sample.
Figure \ref{fig:ccf_2d} presents the CCFs for the extended-EMPG and the comparison samples.

We fit the CCFs with a simple power-law model with two parameters, $A_\omega$ and $\beta$,
\begin{equation}
    \omega_{\rm p}(r_{\rm p}) = A_\omega {r_{\rm p}}^{-\beta}. \label{eq:ccf}
\end{equation}
We perform the fitting 
using the maximum likelihood estimation
with all data points
as well as the data points only in the large scale of $r_{\rm p} > 2$ Mpc 
that are free from
the one-halo term (i.e., galaxy pairs from the same halos) including the contribution from cluster-sized halos of $\sim1$ Mpc.
We present the best-fit power-law models in Figure \ref{fig:ccf_2d} and the best-fit parameters in Table \ref{tab:ccf}.

\begin{figure}
    \epsscale{1}
    \plotone{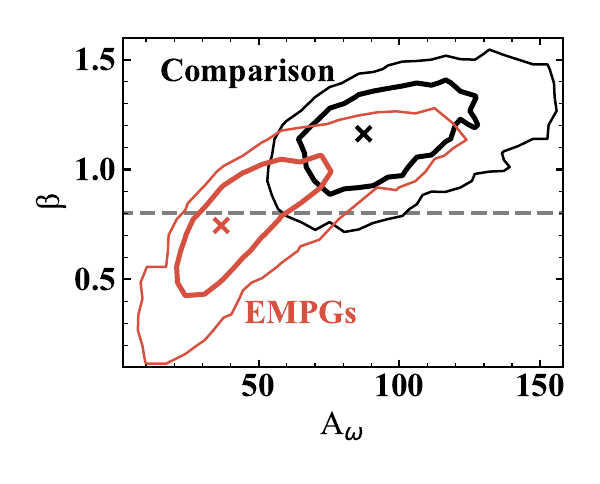}
    \caption{Error contours of the CCF parameters
    for the all data points of the extended-EMPG and comparison samples. 
    The thick and thin red (black) lines indicate the 68\% and 95\% confidence levels for the CCFs of the extended-EMPGs (comparison) sample, respectively. The red (black) cross displays the best-fit parameters for the CCF of the extended-EMPGs (comparison) sample. The dashed gray line denotes the fiducial value of $\beta=0.8$.
    \label{fig:ccf_err}}
\end{figure}

\begin{figure}
    \epsscale{1}
    \plotone{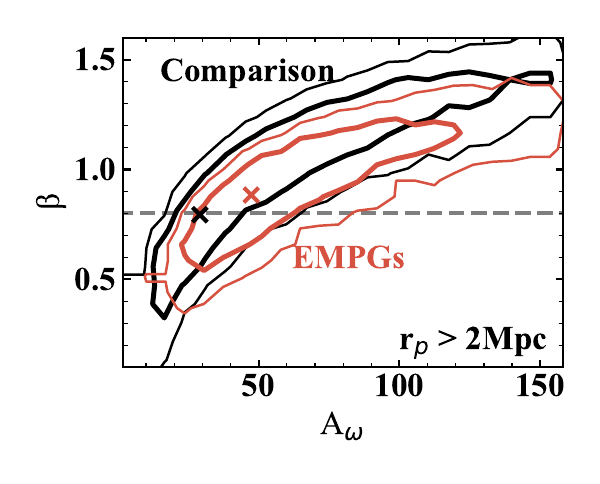}
    \caption{Same as Figure \ref{fig:ccf_err}, but for the data points only in the large scale of $r_{\rm p} > 2$ Mpc.
    \label{fig:ccf_err_L}}
\end{figure}

We derive the parameter errors using Markov chain Monte Carlo (MCMC) parameter estimation technique. 
Figures \ref{fig:ccf_err} and \ref{fig:ccf_err_L} are the error contours of the parameters from our MCMC run for all the data points and the data points in the large scale, respectively.
In Figure \ref{fig:ccf_err} for all the data points,
we find no significant ($>1\sigma$) difference between the error contours of the two samples of the extended EMPG and the comparison samples. In Figure \ref{fig:ccf_err_L} for the data points in the large scale, the contours of the two samples are even similar.
As it is difficult to constrain these two parameters with the limited statistics of the CCFs for the two samples,
we fix the $\beta$ value to the fiducial value, $\beta = 0.8$,
that is adopted in previous clustering analyses \citep[e.g.,][]{Ouchi2001,Ouchi2004b,Ouchi2010,Foucaud2010,Harikane2016}.
The best-fit values of $A_\omega$ and their 2$\sigma$ uncertainties of the EMPG (comparison) sample are $40_{-11}^{+37}$ ($75_{-20}^{+52}$) and $40_{-12}^{+46}$ ($30_{-9}^{+39}$), for the cases of all the data points and the data points in the large scale, respectively.
Even if the $\beta$ value is fixed to the fiducial value, 
the differences of $A_\omega$ 
in the two samples is
within the $<2\sigma$ level
for the both cases.
We note that the error estimation 
with Equation \ref{eq:poisson} can underestimate errors because this error estimation do not capture the cosmic variance introduced by large-scale structures. However, making the error evaluation more plausible would not change our conclusion that there is no difference within errors.

\begin{table}
\caption{Best-fit parameters of CCFs.}
\label{tab:ccf}
\begin{center}
\begin{tabular}{cccc}
\hline\hline
Sample & $r_{\rm p}$ (Mpc) & $A_\omega$ & $\beta$ \\ 
(1) & (2) & (3) & (4) \\ \hline
EMPG & 0 -- 40  & $37_{-6}^{+34}$ & $0.75_{-0.17}^{+0.25}$ \\ 
Comparison & 0 -- 40  & $87_{-11}^{+31}$ & $1.16_{-0.18}^{+0.15}$ \\ \hline
EMPG & 2 -- 40  & $47_{-9}^{+59}$ & $0.88_{-0.24}^{+0.28}$ \\ 
Comparison & 2 -- 40  & $29_{-3}^{+83}$ & $0.79_{-0.10}^{+0.53}$ \\ \hline
\end{tabular}
\end{center}
\footnotesize{{\sc Note}---(1) Sample. (2) Data range used to fit the model. (3)--(4) Parameters of the power-law models (Equation \ref{eq:ccf}).}
\end{table}


\section{Chemical Properties}\label{sec:MZR}
We investigate the relation among stellar mass, metallicity, and sSFR of our EMPGs to understand the chemical properties of these galaxies. 
To obtain $M_*$ and SFR of our EMPG sample, 
we exclude galaxies whose 
rest-frame $i$-band photometry
and \Ha\ or \Hb\ fluxes are not available from the literature.
We use 80 EMPGs in total.

We estimate stellar masses in the manner similar to \citetalias{Isobe2021},
using the SED model code, {\sc beagle} \citep{Chevallard2016}.
We run the {\sc beagle} code by changing the parameters of stellar mass $M_*$ and maximum stellar age $t_\mathrm{max}$ in the range of $\log(M_*/M_\odot) = 4.0-10.0$ and $\log(t_\mathrm{max}/\mathrm{yr}) =$ 4.0--8.0, respectively.
We assume the metallicity of $12+\log(\mathrm{O/H})= 8.2$, the constant star-formation history, no dust attenuation $E(B-V)$, and a \citet{Chabrier2003} IMF.
Then we obtain a mass-luminosity ($M_*$ and absolute $i$-band luminosity) relation,
by the linear fitting to the stellar masses and $i$-band luminosities obtained from the resulting SEDs.
We estimate stellar mass of our EMPGs from their $i$-band magnitudes using this relation.

SFRs are calculated based on the dust-corrected H$\alpha$ fluxes,
assuming the \citet{Kennicutt1998} calibration with a \citet{Chabrier2003} IMF.
If the H$\alpha$ line is saturated, we use a dust-corrected H$\beta$ line instead
and use the intrinsic line ratio of H$\alpha$/H$\beta$ = 2.86 for the case B recombination for typical {\sc Hii} regions \citep[i.e., $T_{\rm e} = 10^4$ K and $n_{\rm e} = 10^2\ {\rm cm}^{-3}$;][]{Osterbrock2006}.

\begin{figure}
    \epsscale{1}
    \plotone{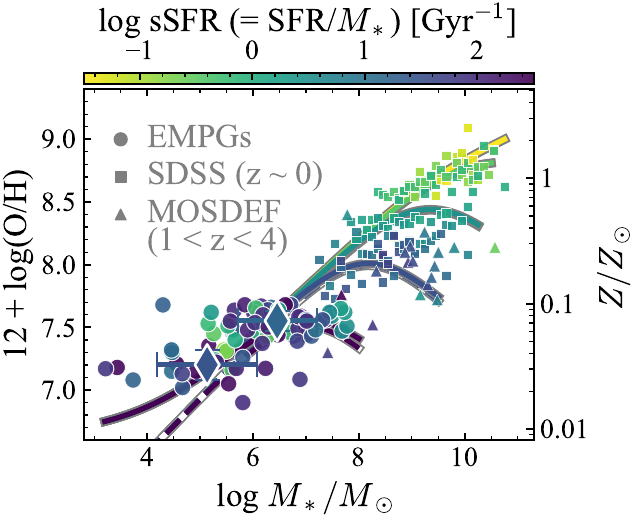}
    \caption{
    Mass-metallicity relation, color-coded by sSFR. 
    The circles and diamonds indicate individual and average values binned by metallicities of our EMPGs, respectively.
    We also show comparison samples of $z\sim0$ galaxies from SDSS \citep[squares;][]{AM13} and high-$z$ ($1<z<4$) galaxies from MOSDEF survey \citep[triangles;][]{Sanders2020}.
    The solid (dashed) lines represent the 
    best-fit model to the $z\sim0$ SDSS galaxies
    at the fixed sSFR of $-1.5$, $-0.5$, 0.5, 1.5, and 2.5 Gyr$^{-1}$,
    with the infall metallicity of $Z_0=$ 0.0034 \Zsun\ ($Z_0=0$).
    \label{fig:fmr}}
\end{figure}

We show the mass-metallicity relation (MZR) of our EMPGs in Figure \ref{fig:fmr}, comparing with the star-forming galaxies at $z\sim0$ \citep{AM13} and $1 < z < 4$ \citep{Sanders2020}.
The data of \citet{AM13}
are the composite spectra that are binned in both $M_*$ and SFR,
while the data of \citet{Sanders2020} are 
individual galaxies. 
All of the data have the direct-method metallicity measurements.

We use analytic chemical evolution models 
to understand whether our EMPGs satisfy the balance between inflow and outflow, 
just like typical local galaxies.
We model metallicities of $z \sim 0$ SDSS galaxies by employing the ideal gas-regulator model with steady inflow and outflow \citep{Lilly2013}. 
In this model, the equilibrium value of metallicity $Z_\mathrm{eq}$ is expressed as

\footnotesize
\begin{eqnarray}
     &Z_\mathrm{eq}& = Z_0 \nonumber\\
     &+& \frac{y(1-R)}{(1-R)+\lambda+\varepsilon^{-1}\left\{(1-R)(1+\beta-b)\mathrm{sSFR}-\frac{1.2}{t}\right\} } \label{eq:zeq}
\end{eqnarray}
\normalsize
\noindent
where $\lambda$ is the mass-loading factor, $R=0.4$ is a return fraction, $\varepsilon$ is the star formation efficiency, $\beta=-0.1$ is the logarithmic slope of the mass dependence of the sSFR, and $t$ is set to 13.8 Gyr. As in \citet{Lilly2013}, we assume that both $\varepsilon$ and $\lambda$ are represented by power laws in the stellar mass of the galaxy, i.e.,
\begin{eqnarray}
    \lambda = \lambda_{10}{m_{10}}^a \label{eq:mlf}\\
    \varepsilon = \varepsilon_{10}{m_{10}}^b \label{eq:sfe}
\end{eqnarray}
where $m_{10}$ is the stellar mass in units of $10^{10}\ M_\odot$. 
We fit using Equation (\ref{eq:zeq}) to $Z$($M_*$, SFR) data 
for SDSS galaxies at $z\sim0$
obtained by \citet{AM13}.
We have five free parameters, $\varepsilon_{10}$, $\lambda_{10}$, $a$, $b$, and $y$ in Equation (\ref{eq:zeq}). We assume the infall metallicity $Z_0$ to be 0.
We fit to the data using the {\tt curve\_fit} function from the Python package scipy. 
The best-fit parameters are given in Table \ref{tab:mzr}.
The dashed lines in Figure \ref{fig:fmr} represent the obtained model of $Z$($M_*$, sSFR)
at the fixed sSFR of $-1.5$, $-0.5$, 0.5, 1.5, and 2.5 Gyr$^{-1}$.
As this model does not explain a metallicity floor, 
we introduce a non-zero infall metallicity 
to consider IGM metal enrichment.
Without changing the other parameters, 
the infall metallicity of $Z_0=0.0034$ \Zsun, instead of 0,
minimize the mean squared error between 
the $Z(M_*$, SFR) of the model and that of the $z\sim0$ SDSS galaxies.
The model with $Z_0=0.0034$ \Zsun\ is shown by the solid lines in the Figure \ref{fig:fmr} and used in the following discussions.

We divide our EMPGs into two groups by their metallicities with \Oabundance\ $=$ 6.69--7.38 and \Oabundance\ $=$ 7.38--7.69.
The mean and standard deviation of the metallicity, stellar mass, and SFR in each group are
\Oabundance\ $=7.20\pm0.12$ (7.56 $\pm$ 0.08), $\log (M_*/M_\odot) = 5.13\pm0.94$ (6.45 $\pm$ 0.75), and log SFR $=$ $-2.44\pm1.33$ ($-1.35\pm2.24$) for the group with \Oabundance\ $=$ 6.69--7.38 (7.38--7.69).
The metallicity of the model at the same stellar mass and SFR, $Z(M_*$, SFR), are 
$Z (\log (M_*/M_\odot)=5.13, \mathrm{SFR}=-2.44) =$ 7.15 and 
$Z (\log (M_*/M_\odot)=6.45, \mathrm{SFR}=-1.35) =$ 7.60.
We find that the prediction on the low-metallicity regime of the model, which is fitted to the galaxies with higher metallicities, agree with the mean values of the EMPGs within the scatter.
This result suggests that on average, our EMPGs are in equilibrium i.e., satisfying the balance between gas inflow and outflow.

\begin{table*}
\caption{Best-fit parameters of Equation \ref{eq:zeq}.}
\label{tab:mzr}
\begin{center}
\begin{tabular}{ccccc}
\hline\hline
log $y$ & $\lambda_{10}$ & a & ${\varepsilon_{10}}^{-1}$(Gyr) & $b$ \\ 
(1) & (2) & (3) & (4) & (5) \\ \hline
9.09 $\pm$ 0.17 & 0.55 $\pm$ 0.42 & -0.44 $\pm$ 0.071 & 0.85 $\pm$ 0.47 & -0.43 $\pm$ 0.07 \\ \hline
\end{tabular}
\end{center}
\footnotesize{{\sc Note}---
(1) Yield in units of $12-\log(\rm{O/H})$.
(2) Coefficient of the mass-loading factor in Equation \ref{eq:mlf}.
(3) Exponent of the mass-loading factor in Equation \ref{eq:mlf}.
(4) Coefficient of the star formation efficiency in Equation \ref{eq:sfe}.
(5) Exponent of the star formation efficiency in Equation \ref{eq:sfe}.}
\end{table*}

\section{Extending the Technique to the High-$z$ EMPG Search}\label{sec:jwst}
Although the possible metallicity floor of 0.01 \Zsun\ exists in the local galaxies (Section \ref{sec:intro}),
 galaxies with $Z < 0.01$ \Zsun\ are more likely to exist at higher redshift 
 as the mean metallicity of the Universe is predicted to be decreasing with increasing redshift \citep{Madau2014}.
To search for high-$z$ EMPGs, we need deep infrared data that cover the rest-frame optical light of high-$z$ galaxies. 
It is now possible
to extend the successful color excess technique (Section \ref{sec:searching}) to a high-$z$ EMPG search,
thanks to unprecedentedly deep and high quality imaging data covering 1--5$\mu$m taken by JWST/NIRCam.

\subsection{SED Model}
\begin{figure*}
    \epsscale{1}
    \plottwo{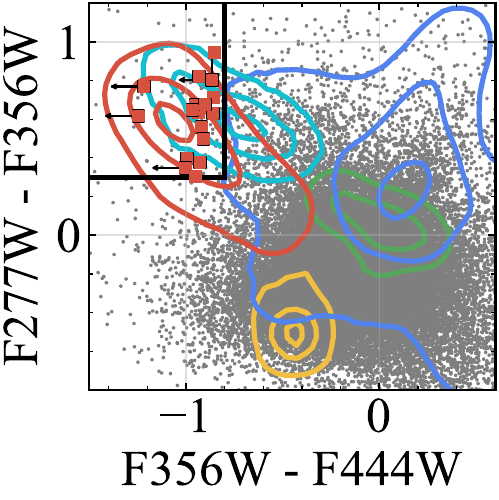}{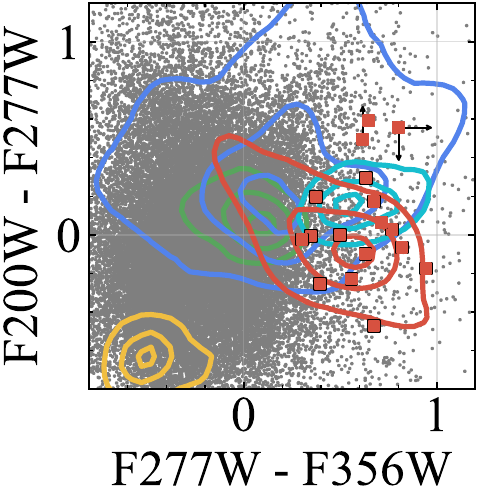}
    \caption{
    Color-color diagrams of F277W$-$F356W vs. F356W$-$F444W (left) and F200$-$F277W vs. F277W$-$F356W (right).
    The gray points show NIRCam sources which are detected at $>3\sigma$ in all three bands of the each panel.
    The red contours illustrate the distribution of EMPGs at $z\simeq$ 3.8--4.8 with metallicities of $Z=$ 0--0.05 \Zsun\ based on the SED model,
    while the other colors correspond to other populations of 
    metal rich galaxies at $z\simeq$ 3--8 (blue), 
    QSOs at $z\simeq$ 3--8 (green),
    and galactic stars (yellow).
    The cyan contours show \OIII\ emitters at $z\simeq$ 6.6--6.7.
    The black lines in the left panel are our color criteria to select the EMPG candidates at $z\sim$ 4--5.
    The selected 17 candidates are marked with the red squares.
    \label{fig:jwst_cc1}}
\end{figure*}

To investigate 
if we can use NIRCam photometric data 
to isolate EMPGs from other type objects of metal-rich galaxies, QSOs, and stars,
we use SED models of EMPGs, metal-rich galaxies, QSOs, and stars used in Section \ref{sec:searching}.
For the EMPG models, 
in addition to those used in Section \ref{sec:searching}, 
we employ the SED models of PopIII \citep{NM22}
with metallicities of $Z = 0$,
stellar ages of 6.0 $\le$ log(age/yr) $\le$ 7.8 with a step of 0.2,
ionization parameters of $-2.0 \le \log U \le -0.5$ with a step of 0.5,
dust attenuations of $E(B-V) =$ [0, 0.01, 0.02, 0.05],
and a \citet{Salpeter1955} IMF with mass ranges of 1--100 \Msun, 1--500 \Msun\ and 50--500 \Msun \citep{Schaerer2003}.
We also use non-PopIII EMPG models of \citet{NM22}
with metallicities of $Z =$ [1/1400, 1/140, 0.01, 0.02, 0.05] \Zsun,
stellar ages of 6.0 $<$ log(age/yr) $<$ 7.8 with a step of 0.2,
ionization parameters of $-3.0 \le \log U \le -1.0$ with a step of 0.5,
dust attenuations of $E(B-V) =$ [0., 0.01, 0.02, 0.05],
and a \citet{Kroupa2001} IMF with upper mass limits of 100 \Msun\ and 300 \Msun.
We obtain colors of NIRCam bands with the model SEDs,
which include random noise corresponding to the observational errors,
in similar procedures as described in Section \ref{sec:ml_catalog}.
Figure \ref{fig:jwst_cc1} shows color-color diagrams of the bands of F200W, F277W, F356W, and F444W.
The red contours in Figure \ref{fig:jwst_cc1} represent 
EMPG models at $z\simeq$ 3.8--4.8 (i.e., $z\sim$ 4--5)
whose \Hb$+$\OIII\ and \Ha\
fall in F277W and F356W, respectively, 
while continuum with no strong lines is sampled by F200W and F444W.
As shown in Figure \ref{fig:jwst_cc1}, $z\sim 5$ EMPG candidates can be selected by the color-excess technique. 

\subsection{JWST Data}
We analyze four JWST/NIRCam datasets 
taken by the Early Release Observations 
\citep[ERO;][]{Pontoppidan2022}
and Early Release Science (ERS) programs: 
ERO SMACS J0723, 
ERO Stephan’s Quintet, 
ERS Cosmic Evolution Early Release Science 
\citep[CEERS;][]{Finkelstein2017,Finkelstein2022c},
and ERS GLASS \citep{Treu2017,Treu2022}. 
The total area used in our analysis is { 87.4} arcmin$^2$.
The 5$\sigma$ limiting magnitudes in the F356W band 
ranges between 28.9 and 29.9 mag.
We use the photometric catalogs constructed by \citet{Harikane2022c}.
The data reduction and photometry are described in \citet{Harikane2022c}.
Following \citet{Harikane2022c},
we measure the object colors with
the 0\farcs3-diameter aperture magnitude in PSF-matched images.
For simplicity,
we use {\sc SExtactor} {\tt MAG\_AUTO} as a total magnitude.

\subsection{Candidate Selection}
We select $z\sim$ 4--5 EMPG candidates based on broadband colors
by adopting the following color criteria:
\begin{eqnarray}
    \rm{F356W} - \rm{F444W} < -0.8 \\
    \rm{F277W} - \rm{F356W} > 0.3
\end{eqnarray}
In calculation of colors, a magnitude fainter than a 3$\sigma$ limiting magnitude is replaced by the 3$\sigma$ limiting magnitude.
We further require $> 3\sigma$ detections 
within 0\farcs2-diameter circular apertures
in the F356W band.
Note that we do not rely on a machine-learning technique
because the number of JWST sources that we need to visually inspect 
is small even if we adopt simple color cuts
due to the smaller size of the JWST source catalogs
than that of SDSS.
In the future with more data, a machine-learning technique would be useful for the high-$z$ EMPG search as well.

\begin{figure}
    \epsscale{1}
    \includegraphics[width=\linewidth]{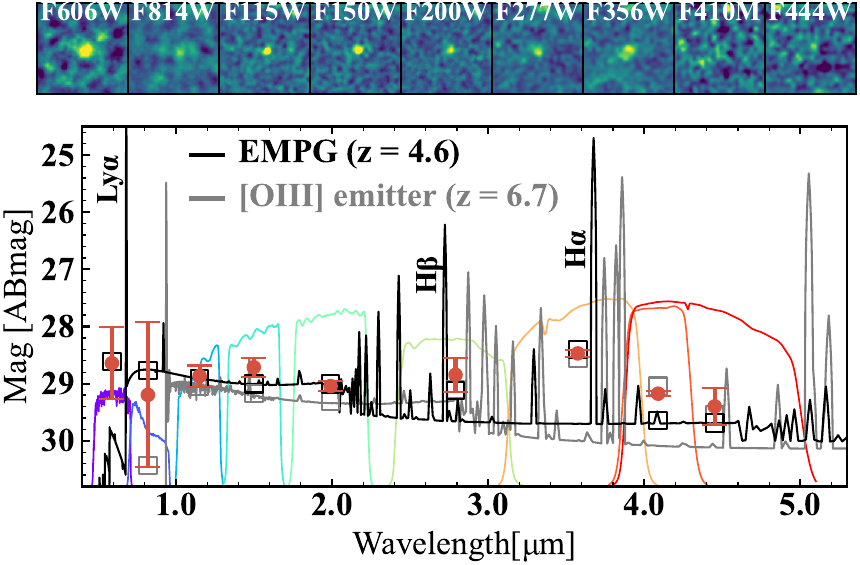}
    \caption{
    SED of one of the JWST EMPG photometric candidates 
    (red circles with error bars) at R.A. = 14h 19m 31.699s, Dec. $= +$52d 50m 39.660s (J2000).
    The candidates present enhanced F356W and F277W photometry, 
    which are explained by the intense \Ha\ and \Hb\
    and modest \OIII\ lines, 
    as indicated by an EMPG model SED at $z\simeq4.6$ (black line and squares). 
    Note that F606W photometry of the EMPG model is boosted by strong \Lya\ emission.
    the gray line and squares show SED of an \OIII\ emitter model spectrum at $z\simeq6.7$,
    which mimic the colors of $z\sim$ 4--5 EMPGs.
    The \OIII\ emitter model SED cannot explain the F606W photometry of the candidate.
    Colored lines show NIRCam and HST filter throughputs of F606W, F814W, F115W, F150W, F200W, F277W, F356W, F410M, and F444W, from left to right.
    The upper nine panels show NIRCam and HST images
    ($1\farcs5 \times 1\farcs5$).
    \label{fig:jwst_sed1}}
\end{figure}

We visually inspect images of the candidates
to remove spurious sources 
and sources affected by bright neighbors and diffraction spikes. 
We then compare SEDs of the candidates, including bluer bands (F200W, F150W, F115W, and F090W if available), with an EMPG spectral model at $z\simeq4.6$ with $Z = 0$, $\log U\simeq -2.0$, a stellar age of $\sim 10^6$ yr, and $E(B-V)\simeq 0.01$ (as shown in Figure \ref{fig:jwst_sed1}). The EMPG model is normalized by average magnitudes in the F277W, F356W, and F444W bands. Additionally, we use multi-band images obtained by the Hubble Space Telescope (HST) ACS in the SMACS J0723 and CEERS fields, which are available on the websites of RELICS \citep{Coe2019}\footnote{\url{https://archive.stsci.edu/prepds/relics}}
and CEERS\footnote{\url{https://ceers.github.io/releases.html}}, respectively. We perform photometry for the candidates on the HST images in the F606W and F814W bands of the CEERS fields, and the F435W, F606W, and F814W bands of the SMACS J0723 field. We remove sources whose SEDs are not consistent with the EMPG model.

\begin{figure}
    \epsscale{1}
    \includegraphics[width=\linewidth]{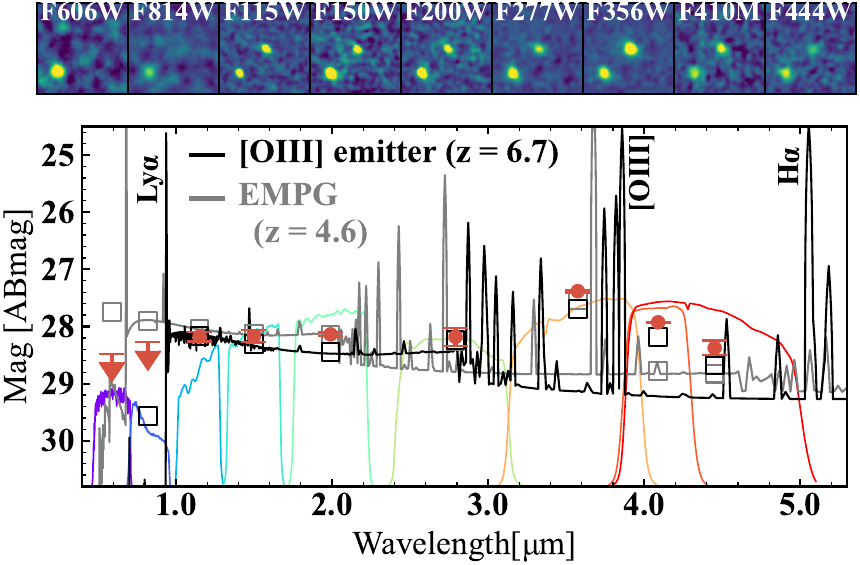}
    \caption{
    Same as Figure \ref{fig:jwst_sed1}, but for a possible \OIII\ emitter at R.A. = 14h 19m 26.971s, Dec. $= +$52d 48m 37.980s (J2000).
    The black (gray) line and squares show SED of an $z\simeq6.7$ \OIII\ emitter ($z\simeq$ 4.6 EMPG) model spectrum.
    \label{fig:jwst_sed2}}
\end{figure}

One of the main sources removed from the candidate list is a \OIII\ emitter at $z\simeq$ 6.6--6.7 because it shows strong \OIII\ emission in F356W and no strong emission in F444W. The black (gray) line in Figure \ref{fig:jwst_sed2} (Figure \ref{fig:jwst_sed1}) represents an \OIII\ emitter spectral model at $z\simeq6.7$ with $Z = 0.2\ Z_\odot$, $\log U\simeq -2.5$, stellar age of $\sim 10^{6.2}$ yr, and $E(B-V)\simeq 0.01$. While the redshift range that the \OIII\ emitter model satisfies our color criteria is narrower ($\Delta z\sim0.1$) than EMPGs ($\Delta z\sim1$), strong \OIII\ emission-line emitters are abundant at $z\gtrsim6$ \citep[e.g.,][]{Matthee2022} and can be contaminants in our sample.
We distinguish between $z\sim$ 4--5 EMPGs and $z\sim6.7$ \OIII\ emitters based on the Lyman break at $\sim$ { 600--700} nm and $\sim$ { 930} nm, respectively. The expected magnitude in the F090W band of the \OIII\ emitter model spectrum is 0.5 mag fainter than that of the EMPG model spectrum due to the Lyman break. Therefore, we remove sources with drops in F090W. 
We also use HST multi-band images covering wavelengths shorter { than $\sim$ 930 nm to remove $z\sim6.7$ \OIII\ emitters.} We remove objects that are not detected in these HST bands from the candidate list because EMPGs at $z\sim$ 4--5 would be brighter than the 2$\sigma$ limiting magnitudes according to our model predictions. Figure \ref{fig:jwst_sed2} illustrates one of the potential \OIII\ emitters that satisfies our color criteria but is removed from the candidate list.

Finally, we find 17 EMPG candidates that exhibit characteristic colors similar to $z\sim$ 4--5 EMPGs. Figure \ref{fig:jwst_sed1} shows the EMPG spectral model and SED for one of the 17 candidates. The signal-to-noise ratio within 0\farcs2-diameter circular apertures in the F356W band for all 17 candidates is greater than 4. Among the 17 candidates, one is marginally detected in the F606W (2.6$\sigma$) and F814W (1.5$\sigma$) bands, which are consistent with the SED of the EMPG model (Figure \ref{fig:jwst_sed1}). Of the remaining 16 candidates, eight have F090W photometry, showing flat colors in the F090W, F150W, and F200W bands. However, we cannot place constraints on the remaining 8 candidates using either F090W photometry or HST images because of the shallow detection limits of the HST images, which are at around $\sim$ 26--29 mag, and the lack of F090W photometry.
While a spectroscopic confirmation is necessary to further investigate the contamination, we retain the 17 candidates in the $z\sim$ 4--5 EMPG sample in this paper based on the NIRCam and HST photometry.

\subsection{Line Ratio Estimation}
\begin{figure}
    \includegraphics[width=\linewidth]{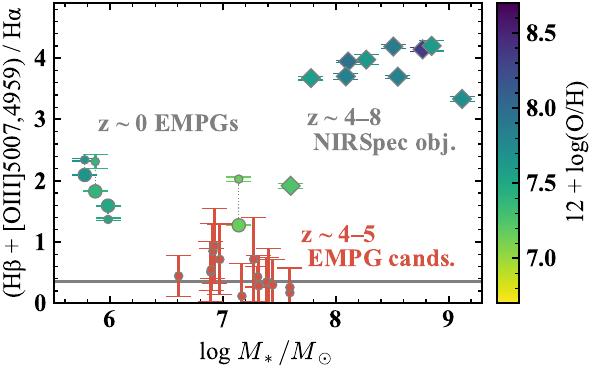}
    \caption{
    Strong line ratio of (\Hb$+$\OIII$\lambda\lambda5007,4959$)$/$\Ha\ as a function of stellar mass.
    The $z\sim$ 4--5 EMPG candidates are shown with the red points with errors of the line ratio.
    The green points and circles denotes 
    line ratios of the photometric estimates and the spectroscopic results, respectively, for $z\sim0$ EMPGs from this work (Section \ref{sec:searching}) and the literature \citep{Izotov1998}.
    The diamonds are the 
    { $z \simeq$ 4--8}
    galaxies identified by NIRSpec spectroscopy { \citep{Nakajima2023}}.
    The $z\sim0$ EMPGs and { NIRSpec objects} are color-coded according to metallicities based on the direct $T_\mathrm{e}$ method.
    The gray horizontal line represents the theoretical lower limit with negligible metal emission of \OIII, i.e., H$\beta/$H$\alpha$,
    for the case B recombination at $T_{\rm e} = 10^4$ K and $n_{\rm e} = 10^2\ {\rm cm}^{-3}$.
    \label{fig:jwst_ratio}}
\end{figure}

We estimate 
\Hb$+$\OIII\ and \Ha\ line fluxes of the candidates
with F277W and F356W, respectively, 
with the continuum estimated with F444W,
in the same manner as \citetalias{Isobe2021}.
The candidates show large color excesses in the F356W band of F356W $-$ F444W $\sim-1$, 
which are mainly caused by strong \Ha\ lines.
We regard the observed F444W-band flux density $f_\mathrm{F444W}$ as tracers of the stellar continuum with a negligible effect of strong emission lines.
We also assume that the flux densities per unit frequency of stellar continuum is constant 
{ over the wavelength range of \Hb, \OIII, and \Ha.}
We then calculate the \Ha\ flux $f_\mathrm{H\alpha}$ as follows: 
\begin{equation}
 f_\mathrm{H\alpha} = (f_\mathrm{F356W} - f_\mathrm{F444W}) \times \frac{c}{\lambda_\mathrm{F356W}} \times \Delta\lambda_\mathrm{F356W},
\end{equation}
where $f_\mathrm{F356W}$, $\lambda_\mathrm{F356W}$, and $\Delta\lambda_\mathrm{F356W}$
represent the observed F356W-band flux density, the central wavelength of the F356W-band filter, and the width of the F356W-band filter, respectively. 
Errors of the $f_\mathrm{H\alpha}$ is calculated from photometric errors.
We estimate the sum of \Hb$+$\OIII\ fluxes and their errors in the same manner as the one of the \Ha\ fluxes.
Figure \ref{fig:jwst_ratio} presents (\Hb$+$\OIII)/\Ha\ ratios
as a function of stellar mass derived by the mass-luminosity relation used in Section \ref{sec:MZR}.

For comparison, 
we photometrically estimate line ratios of (\Hb$+$\OIII)/\Ha\ for $z\sim0$ EMPGs,
including the three newly identified EMPGs (Section \ref{sec:searching}) and I Zw18 NW \citep{Izotov1998}, one of the well-known EMPGs,
in the same manner as the JWST EMPG photometric candidates. 
The \Hb$+$\OIII\ and \Ha\ line fluxes are estimated
with SDSS $g$- and $r$-band photometry, respectively,
with the continuum estimated with $i$-band photometry.
Errors of the estimated line fluxes are calculated from photometric errors. 
Compared to
the spectroscopic results of the
line ratios for these three EMPGs (Section \ref{sec:mage-obs})
and I Zw18 NW \citep{Izotov1998},
the spectroscopic results agree with the photometric estimates within 0.05 -- 0.20 dex.
We confirm that the method of the photometric estimates
does not produce the small (\Hb$+$\OIII)/\Ha\ ratios by estimation systemtics.
We can also compare our results with those obtained from high-$z$ galaxies in a similar redshift range, whose optical emission lines have been spectroscopically identified using JWST/NIRSpec. We present 10 galaxies at $z=$ 4--8.5 in Figure \ref{fig:jwst_ratio} that are found from the three major public spectroscopy programs of ERO, GLASS, and CEERS with reliable metallicity determinations based on \OIII$\lambda$4363 \citep{Nakajima2023}.
We estimate the \Ha\ flux from the \Hb\ flux for the NIRSpec objects
assuming the line ratio of H$\alpha$/H$\beta$ = 2.86 
for the case B recombination at $T_{\rm e} = 10^4$ K and $n_{\rm e} = 10^2\ {\rm cm}^{-3}$.

Comparing the $z\sim$ 4--5 EMPG candidates with the $z\sim0$ EMPGs and the { $z\simeq$ 4--8} NIRSpec galaxies in Figure \ref{fig:jwst_ratio}, we find that the $z\sim$ 4--5 EMPG candidates exhibit exceptionally small (\Hb$+$\OIII)/\Ha\ ratios.
Interestingly, 14 out of the 17 candidates have 
very low (\Hb$+$\OIII)/\Ha\ ratios
consistent with negligible \OIII\ emission
(i.e., \Hb/\Ha; gray horizontal line in Figure \ref{fig:jwst_ratio}), 
indicative of PopIII-like galaxies. 
Although dust extinctions are not corrected for the ratios
of JWST EMPG candidates
as well as $z\sim0$ EMPGs and NIRSpec objects,
the ratios increase only by 0.05 dex
by the typical extinction values for local EMPGs of $E(B-V)=0.1$, 
which is much smaller than the error bars ($\sim0.4$ dex).

\section{Discussion}\label{sec:discussion}
\subsection{Metallicity Floor}
Although we search for EMPGs with SED models including galaxies with $Z <$ 0.01 \Zsun\ { (i.e., HMPGs)},
we have not discovered HMPGs (Section \ref{sec:intro}).
We discuss three possibilities to explain the absence of HMPGs
\citep[see also][]{Morales-Luis2011,McQuinn2020}.

The first possibility is faintness of HMPGs.
HMPGs are expected to have stellar masses of $M_*\sim10^{4.5}$ \Msun\
based on the MZR of our model obtained in Section \ref{sec:MZR} that predicts
the stellar mass of $M_*\sim10^{4.5}$ \Msun\ at sSFR $\sim$ 1 and $Z \sim$ 0.01 \Zsun.
From the mass-luminosity relation used in Section \ref{sec:MZR},
a galaxy with $M_*\sim10^{4.5}$\Msun\ has a $i$-band absolute magnitude of $-10.3$,
which is difficult to detect with SDSS survey with the $i$-band limiting magnitude of 21.3
because it corresponds to a $i$-band apparent magnitude of $m_i\simeq25$ (23) at $z=0.03$ (0.01).
Deep imaging surveys such as HSC survey,
whose $i$-band limiting magnitude is $m_i \simeq 26$,
would be useful to search for HMPGs.
However, we should note that no HMPG has been found in previous EMPG searches even with the HSC data
\citepalias{Kojima2020,Isobe2022a,Nakajima2022}.

The second possibility is that 
metal-poor gas in a galaxy may be enriched to $> 0.01$ \Zsun\
as soon as star formation begins.
Considering short lifetimes of massive stars
($\sim$ a few Myr)
and the typical age of EMPGs ($\sim50$ Myr; \citetalias{Isobe2022a}),
the timescale for enrichment to $>$ 0.01 \Zsun\ can be short,
making observable HMPGs rare.
We note that although the youngest EMPGs observed to date have sSFR of $\sim300$ Gyr$^{-1}$, which corresponds to the mass-doubling time (1/sSFR) of $\sim 3$ Myr, none of them have $Z < 0.01$ \Zsun.

The third possibility is that
the part of the IGM accreting on EMPGs is
already enriched to $>$ 0.01 \Zsun\ 
by past star formation.
The mean metallicity of the Universe 
predicted from cosmic star formation density measurements
increases as decreasing redshift 
and reaches $\sim$ 0.1 \Zsun\ at $z = 0$ \citep{Madau2014}.
HMPG as well as IGM with metallicity below 0.01 \Zsun\ could be too rare to be found by current surveys in the local universe.
Going to higher redshift, the mean metallicity of the IGM as probed by the \Lya\ forest at $z\sim$ 2--3 has been estimated to be between 0.001--0.01 \Zsun\
\citep{Schaye2003,Simcoe2004,Aguirre2008}.
This is in agreement with our finding, in Section \ref{sec:MZR}, that the non-zero infall metallicity best reproducing the shape of the metallicity floor of the MZR, as measured for the local SDSS galaxies, is as low as 0.0034 \Zsun. 
Also, the measured and predicted metallicities at $z\gtrsim4$
are $\lesssim$ 0.001 \Zsun\ 
\citep{Simcoe2011,Madau2014}.
This suggests that true HMPGs may be found at $z\sim$ 4--5, like the candidates presented in Section \ref{sec:jwst}.

In this study, we find no HMPGs in the SDSS imaging data. 
Although we cannot conclude which of the three possibilities is more plausible, we estimate the upper limit of the number density of HMPGs from this result.
From the discussion above, we assume that HMPGs have the $i$-band absolute magnitudes of $-10.3$.
Since we select EMPG candidates using the SDSS imaging data with $i$-band magnitudes below 21.3 (Section \ref{sec:photo-selection}), 
we can search for HMPGs up to a luminosity distance of $d_L\simeq$ 21 Mpc (corresponding to $z\simeq$0.005). 
We find 104 EMPG candidates with emission lines 
from the SDSS imaging data with the total area of 14,555 deg$^2$, 
of which the ten candidates were confirmed by deep spectroscopic observations. 
Thus, assuming that we have searched for $\sim10\%$ (10/104) of the total area of the SDSS imaging data (14,555 deg$^2$ $\times\ 10/104 = 1,413$ deg$^2$), we have examined the volume of $1.3\times10^3$ Mpc$^3$.
No HMPGs in that volume means that the upper limit of number density of HMPGs is approximately $7.7\times10^{-4}$ Mpc$^{-3}$. This is about four times smaller than the the number density of EMPGs, $(3.4\pm0.9)\times10^{-3}$ Mpc$^{-3}$, calculated by \citetalias{SA16}. 
We also estimate the number density of $z\sim$ 4--5 EMPGs from the result of Section \ref{sec:jwst}. We select 17 EMPG candidates in the redshift range of $3.8 < z < 4.8$ using the JWST/NIRCam data with a survey area of 87.4 arcmin$^2$. From the volume corresponding to this redshift range and the survey area, and assuming that all of these candidates are genuine, the number density of $z\sim$ 4--5 EMPGs is estimated to be $6.7\times 10^{-5}\ \mathrm{Mpc^{-3}}$, which is lower than the upper limit of the number density of local HMPGs  mentioned earlier.

\subsection{Physical Origin of EMPGs}
As mentioned in Section \ref{sec:intro}, there are two possible scenarios that explain the physical origins of local EMPGs: the episodic and first star-formation scenarios. 
In the episodic star formation scenario, EMPGs have intermittent star-formation histories, where stochastic accretion events of metal-poor gas occur repeatedly on cosmological timescales, causing star formation. As a result, EMPGs have high sSFRs and low gas-phase metallicities. 
In contrast, in the first star-formation scenario, EMPGs are experiencing their first starburst with very metal-poor gas similar to the primordial gas.

We compare the CCF of our EMPG sample with that of the comparison sample, which is composed of galaxies with similar stellar masses to those of EMPGs but higher metallicities, 
and find no significant difference between the CCFs (Section \ref{sec:clustering}).
In other words, even though EMPGs and galaxies in the comparison sample tend to reside in halos with similar mass,
only EMPGs have the metal-poor gas.
This is consistent with the episodic star-formation scenario: stochastic metal-poor gas accretions on metal-enriched galaxies result in low metallicities of EMPGs.
On the other hand, for the first star-formation to occur in the current universe, star formation need to have been suppressed until now by, for example, the ultraviolet (UV) background radiation. However, whether the gas is evaporated by heating due to the UV background radiation is mainly determined by the halo mass and redshift \citep{Hoeft2006}. Therefore, it is theoretically unlikely that only some halos with the same masses have suppressed star formation until now.

Moreover, in Section \ref{sec:MZR}, we show that the MZR of EMPGs is comparable to the that of the model which is fitted to the local SDSS galaxies, which satisfies the balance between inflow and outflow. Although this result does not rule out the first star-formation scenario,
the episodic star-formation scenario is more likely 
because our EMPGs can be explained in the same way as for the local SDSS galaxies with the equilibrium model.
In summary, two findings of this study suggest that EMPGs can be explained, on average, by the episodic star-formation scenario.

This picture is consistent with previous studies
from the aspects of morphologies and metal's spatial inhomogeneities
(e.g., \citealt{SA15}; \citetalias{Isobe2021}; Nakajima et al. in prep)
although in terms of chemical abundances, the physical origin of EMPGs are still controversial (see Section \ref{sec:intro}).

\section{Summary}\label{sec:summary}
We search for EMPG candidates, 
especially for galaxies below the observed metallicity floor of 
$\sim 0.01$ \Zsun, 
using broadband-color excesses.
To select local EMPG candidates, 
we develop our selection method 
using a machine learning technique 
and
apply the selection method to the SDSS imaging data.
We conducted shallow spectroscopic screening observations 
for 134 EMPG candidates 
via strong emission lines 
with the Seimei and Nayuta telescopes
to eliminate possible contamination of stars.
We then performed deep spectroscopy 
for ten EMPG candidates
with the Magellan telescope
to confirm EMPGs via metallicity measurements.
Using the newly identified EMPGs as well as the previously reported EMPRESS and other EMPGs, as compiled in a companion paper,
we statistically investigate the clustering and chemical properties below $Z<0.1$ \Zsun,
and discuss the physical origin of EMPGs.
Our main findings are summarized below. 

\begin{enumerate}
\item 
We identify three out of the ten local EMPG candidates as EMPGs (0.05 -- 0.1 \Zsun)
based on metallicity measurements with faint emission lines including \OIII$\lambda$4363 detected with the Magellan telescope.
The remaining seven candidates are still metal-poor galaxies with 0.1 -- 0.5 \Zsun.

\item 
We find no significant ($>1\sigma$) difference 
between the CCF of our EMPG sample and that of the comparison sample,
which is composed of galaxies with similar stellar masses to those of EMPGs but with higher metallicities.
This is consistent with the episodic star-formation scenario: stochastic metal-poor gas accretions on metal-enriched galaxies result in low metallicities of EMPGs.

\item 
We compare our EMPGs with $z\sim0$ SDSS star-forming galaxies 
on the MZR,
using an analytic chemical evolution model.
We find that on average EMPGs agree with the prediction of the model
that is fitted to the SDSS galaxies,
suggesting our EMPGs satisfy the balance between inflow and outflow as the SDSS galaxies.
This result also supports the episodic star-formation scenario.

\item
We extend our color-excess technique to the $z\sim$ 4--5 EMPG search
using the JWST/NIRCam images taken by ERO and ERS programs.
We select 17 EMPG candidates at $z\sim$ 4--5
with negligible \OIII$\lambda\lambda$4959,5007 emission weaker than the known local EMPG and high-$z$ galaxies, suggesting that some of these candidates may fall in 0--0.01 \Zsun, which potentially break the metallicity floor.

\item 
Since no EMPGs below the known possible metallicity floor of 0.01 \Zsun\ are found at $z\sim0$,
we estimate the upper limit of the number density of galaxies with $Z<$ 0.01 \Zsun\ in the local universe to be approximately $7.7\times10^{-4}$ Mpc$^{-3}$. 

\end{enumerate}

We thank 
K. Shimasaku, 
H. Miyatake, A. Inoue, Y. Takeda, K. Ito, R. Ikeda, S. Chanoul, A. Matsumoto, H. Umeda, S. Aoyama, and S. Kikuta,
for giving us helpful comments.
This paper includes data gathered 
with the 3.8 m Seimei Telescope at the Okayama Observatory,
the 2 m Nayuta Telescope at the Nishi-Harima Astronomical Observatory,
and the 6.5 m Magellan Telescopes located at Las Campanas Observatory.
This paper is supported by World Premier International Research Center Initiative (WPI Initiative), MEXT, Japan, as well as the joint research program of the Institute of Cosmic Ray Research (ICRR), the University of Tokyo. This work is supported by KAKENHI (19H00697, 20H00180, 20K14532, 21H04467, 21H04499, 21J20785, 21K03614, 21K03622, 22H01259, and { 22KJ0157}) Grant-in-Aid for Scientific Research (A) through the Japan Society for the Promotion of Science.
JHK acknowledges the support from the National Research Foundation of Korea (NRF) grants, No. 2020R1A2C3011091 and No. 2021M3F7A1084525, funded by the Korea government (MSIT). 
The English writing in this paper was improved with ChatGPT (OpenAI 2020\footnote{OpenAI (2020), "Introducing ChatGPT", Published on 30 November 2022; Last accessed on 10 April 2023, \url{https://openai.com/blog/chatgpt}}), while no sentences were generated from scratch.

\bibliographystyle{apj}
\bibliography{MAIN.bib}

\end{document}